\documentclass[12pt]{article}

\usepackage{amsmath,amssymb,graphicx} 
\usepackage{pstricks}
\usepackage{cite}
\usepackage[pdfpagemode=FullScreen,bookmarks=true]{hyperref}

\def\beq{\begin{equation}}
\def\eeq#1{\label{#1}\end{equation}}
\def\eeqn{\end{equation}}
\def\beqa{\begin{eqnarray}}
\def\eeqa#1{\label{#1}\end{eqnarray}}
\def\eeqan{\end{eqnarray}}
\def\CR{\nonumber \\ }
\def\leqn#1{(\ref{#1})}

\def\tl{\tilde{t}_1}
\def\th{\tilde{t}_2}
\def\t1{\tilde{t}_1}
\def\tL{\tilde{t}_L}
\def\tR{\tilde{t}_R}

\def\n1{\tilde{\chi}_1^0}
\def\to{\rightarrow}
\def\teff{\theta_{\rm eff}}

\def\met{\mbox{$E{\!\!\!\!}/_{T}$}}

\def\stacksymbols #1#2#3#4{\def\theguybelow{#2}
    \def\vp{\lower#3pt}
    \def\sp{\baselineskip0pt\lineskip#4pt}
    \mathrel{\mathpalette\intermediary#1}}

\def\intermediary#1#2{\vp\vbox{\sp
     \everycr={}\tabskip0pt
     \halign{$\mathsurround0pt#1\hfil##\hfil$\crcr#2\crcr
              \theguybelow\crcr}}}

\begin{document}

 \begin{titlepage}

\vskip.5cm
\begin{center}
{\huge \bf Polarized Tops from Stop Decays at the LHC} \\
\vskip0.4cm
{\huge \bf } 
\vskip.2cm
\end{center}
\vskip1cm

\begin{center}
{\bf Maxim Perelstein and Andreas Weiler} \\
\end{center}
\vskip 8pt

\begin{center}
	{\it Institute for High Energy Phenomenology\\
	Newman Laboratory of Elementary Particle Physics\\
	Cornell University, Ithaca, NY 14853, USA } \\

\vspace*{0.1cm}

{\tt  mp325@cornell.edu,
aw288@cornell.edu}
\end{center}

\vglue 0.3truecm

\begin{abstract}
\vskip 3pt \noindent
In supersymmetric models, scalar top quarks, or stops, generically have 
parity-violating couplings to top quarks. As a result, tops produced in
stop decays should be polarized. In this paper, we will argue that this
effect may be observable at the LHC with realistic integrated luminosities,
provided that one of the stops is copiously produced and can
decay to a top and a neutralino. We define the ``effective'' stop mixing 
angle, which determines the degree of top polarization, and discuss the
prospects for a measurement of this angle at the LHC. If some information 
about the neutralino mixing matrix is available, this measurement can be
used to constrain the mixing angle in the
stop sector, one of the most important ingredients in assessing 
the naturalness of electroweak symmetry breaking in the MSSM.   
\end{abstract}

\end{titlepage}

\section{Introduction}

Supersymmetric (SUSY) extensions of the Standard Model (SM) are one of the 
most 
compelling candidates for physics at the electroweak scale~\cite{HK,Martin}. Among many 
new particles predicted by these models, scalar partners of the top quark,
the stops $\tl$ and $\th$, play a special role. The reason is that the top 
Yukawa 
coupling is by far the strongest coupling of the SM Higgs. The dominant 
quantum contributions to the Higgs potential arise from loops of the top quark 
and its superpartners. These contributions play an important role in the 
physics of electroweak symmetry breaking (EWSB) in this framework. For 
example, in no-scale unified models such as mSUGRA, they are responsible
for triggering EWSB by driving the Higgs mass$^2$ term negative, while all
other scalar masses remain real. More generally, the stop and top loops   
give a sizeable contribution to the parameters determining the EWSB scale
and the Higgs mass spectrum. Thus, if TeV-scale SUSY is realized in
nature and discovered at the LHC, determining the properties of the 
stop quarks will be crucial in understanding the mechanism of EWSB.

In the minimal supersymmetric standard model 
(MSSM), sizable quantum corrections to the physical mass of the lightest 
CP-even Higgs are required to satisfy the LEP-2 lower bound on this mass~\cite{LEPHiggs}. The 
tension between this requirement and naturalness of the EWSB scale is
known as the ``LEP paradox'' or ``little hierarchy problem''~\cite{LEPpar}.
Minimizing this tension requires a certain choice of the stop soft masses and
the trilinear soft A-term~\cite{KN}, pointing to a ``golden region'' in the MSSM
parameter space~\cite{GR}. This region is characterized by fairly light
stops (typically 300-400 GeV for $\tl$ and 500-800 GeV for $\th$), and 
a large mixing angle $\theta_t$ (defined as the rotation angle from the gauge 
eigenbasis $\tL, \tR$ to the mass eigenbasis $\tl, \th$). The shape of the golden region depends only weakly on the non-top sector MSSM parameters (within broad ranges of those parameters), and within the unrestricted MSSM this hypothesis is compatible with other interesting applications of SUSY such as dark matter~\cite{KGF}. If SUSY is 
discovered at the LHC, it would be very interesting to test whether the
golden region hypothesis is realized.

There is a large body of literature on superpartner mass measurements at the
LHC. (For example, Ref.~\cite{MR} studied stop mass determination in a framework similar to the one considered here.) The goal of this paper is to propose observables that are directly sensitive to the mixing angle $\theta_t$. The basic idea is that this angle enters the stop-top-neutralino couplings. In particular, the amount of parity violation in these couplings is determined by $\theta_t$ (along with the neutralino mixing matrix). If the decay $\tilde{t}\to t\chi^0$ is kinematically allowed, this parity violation results in non-zero  polarization of the tops produced in this decay. As is well known, tops decay before they hadronize, so that the top polarization is reflected in the angular distributions of the top decay products~\cite{Kane}. Observing these distributions can therefore provide information on $\theta_t$. In this paper, we will show how such measurements can be performed at the LHC.

The rest of the paper is organized as follows. In Section~\ref{sec:idea}, we discuss the connection between the stop mixing angle and the polarization of tops from $\tilde{t}\to t\chi^0$ decays in more detail, define the effective mixing angle which determines the degree of top polarization, and propose two observables which can be used to measure this effective angle. In Section~\ref{sec:BP}, we perform a parton-level analysis of the feasibility of the proposed measurements at the LHC, for a particular benchmark point (chosen to be compatible with the golden region) in the MSSM parameter space. The analysis includes the leading SM backgrounds, and we discuss a set of selection cuts to suppress these backgrounds to manageable levels. We project the sensitivity of the LHC experiments to the effective mixing angle with 10 fb$^{-1}$ of analyzed data. In Section~\ref{sec:teff}, we briefly discuss some of the issues involved in inferring the actual stop mixing angle $\theta_t$ from the measured effective angle. Finally, we present our conclusions and outline some questions for future study in Section~\ref{sec:conc}. 

\section{Polarized Tops from Stop Decays}
\label{sec:idea}

The stop mass terms in the MSSM Lagrangian have the form
\beq
{\cal L} = (t_L^*, t_R^*)\,M^2\,(t_L, t_R)^T\,, 
\eeq{mstop}
where
\beq
M^2 = \left( \begin{array}{cc} M_L^2+m_t^2+\Delta_u & \sqrt{2} m_t \sin\beta
(A_t-\mu\cot\beta)  \\
\sqrt{2} m_t \sin\beta (A_t-\mu\cot\beta) & M_R^2+m_t^2+\Delta_{\bar{u}} 
\end{array}\right)\,,
\eeq{Mstop}
and
\beq
\Delta_u\,=\,\left(\frac{1}{2}-\frac{2}{3}\sin^2\theta_W \right)
\,\cos2\beta\,m_Z^2\,,~~~~\Delta_{\bar{u}}\,=\,\frac{2}{3}\sin^2\theta_W 
\,\cos2\beta\,m_Z^2\,.
\eeq{deltas}
The mass eigenstates $\tilde{t}_1$ and $\tilde{t}_2$ are superpositions of 
the gauge eigenstates, $\tL$ and $\tR$:
\beqa
\tilde{t}_1 &=& \cos\theta_t \,\tL \,+\,\sin\theta_t\,\tR\,,\CR
\tilde{t}_2 &=& -\sin\theta_t \,\tL \,+\,\cos\theta_t\,\tR\,,
\eeqa{stop_mix}
where the mixing angle is given in terms of the Lagrangian parameters by
\beq
\tan2\theta_t\,=\,\frac{2\sqrt{2} m_t \sin\beta(A_t-\mu\cot\beta)}{M_L^2
-M_R^2 + \Delta_u-\Delta_{\bar{u}}}\,.
\eeq{theta_t}
Since parity is violated in weak interactions, weak couplings of the stops 
depend on the angle $\theta_t$. Of particular interest to us is the 
stop-top-neutralino coupling, since the parity asymmetry in this coupling 
can lead to non-vanishing polarization of top quarks produced in stop decays.
Since the top quark decays before it hadronizes, this polarization is, at least
in principle, observable. The relevant vertex has the form
\beq
g_{\rm eff}^{ij} \tilde{t}_i \tilde{\chi}^0_j \,\left( \cos\teff^{ij} \,P_L 
\,+\, \sin\teff^{ij} \,P_R \right) t\,,
\eeq{vertex}
where $j=1\ldots 4$ labels the neutralino mass eigenstates $\tilde{\chi}_j^0$, 
related to the gauge eigenstates $\tilde{N} = (\tilde{B}, \tilde{W}^3, 
\tilde{H}^0_d, \tilde{H}^0_u)$ by
\beq
\tilde{\chi}_j^0\,=\, \sum_{k=1}^4 N_{jk} \tilde{N}_k\,.
\eeq{Nrot}
The effective mixing angles are given by
\beqa
\tan\teff^{1j} &=&\frac{y_t N_{j4} \cos\theta_t -\frac{2\sqrt{2}}{3}g^\prime 
N_{j1}\sin\theta_t}{\sqrt{2}\left( \frac{g}{2}N_{j2} + 
\frac{g^\prime}{6} N_{j1} \right) \cos\theta_t + y_t N_{j4} \sin\theta_t}\,,\CR
\tan\teff^{2j} &=& \frac{y_t N_{j4} \sin\theta_t +\frac{2\sqrt{2}}{3}g^\prime 
N_{j1}\cos\theta_t}{\sqrt{2}\left( \frac{g}{2}N_{j2} + 
\frac{g^\prime}{6} N_{j1} \right) \sin\theta_t - y_t N_{j4} \cos\theta_t}\,,
\eeqa{thetaeff}
where $y_t = \sqrt{2} m_t/(v\sin\beta)$. The main idea of this paper is that 
we may be able to get an unambiguous and fairly precise experimental 
measurement of one or more of the angles $\teff^{ij}$ at the LHC, by
measuring the polarization of top quarks produced in the decay $\tilde{t}\to
\tilde{\chi}^0 t$. If the neutralino mixing matrix is at least partly
known from other measurements, this information can be used to
extract (or at least constrain) $\theta_t$. This information can in turn be
used, together with the stop eigenmass measurements, to determine the 
stop-sector lagrangian parameters.

At the LHC, stops can be directly pair-produced by strong interactions, 
in the processes 
\beq
pp\to\tilde{t}_i\tilde{t}_i^*. 
\eeq{direct}
Direct production of same-sign stop
pairs is negligible. In addition, there may be a 
sizable sample of stops produced indirectly, namely in decays of other superpartners,
particularly the gluino via $\tilde{g}\to t\tilde{t}$. (A brief discussion of the possibility of top polarization measurements in the gluino sample appeared in Ref.~\cite{Nojiri}.) Those events lead to
more complicated final state topologies in the detector, and vetoing such
topologies can be used to separate the ``direct'' and ``indirect'' stop  
samples. 
We will focus on the direct stop sample in this paper, assuming that 
the contamination from the indirect sample, if present, is negligible. 
This has the advantage of simpler events and more robust predictions, since
the rate and event topologies in the indirect sample depend on many more
MSSM parameters. Some of the analysis techniques described here could be 
applied to the indirect sample as well. 

Once produced, stops will promptly decay. Possible two-body decay modes include
$\tilde{t}\to t\tilde{\chi}^0$, $\tilde{t}\to b\tilde{\chi}^+$, 
$\tilde{t}\to W^+ \tilde{b}$, and $\tilde{t}\to H^+ \tilde{b}$. We are  
interested in the $\tilde{t}\to t\tilde{\chi}^0$ mode, which must be 
kinematically allowed and have a sizeable branching ratio for our analysis to
apply. This decay is followed by $t\to W^+ b$, and the W-boson then decays
either hadronically (about 70\% of events) or leptonically (about 10\% for 
each lepton flavor). Angular distributions of the top decay products are
sensitive to top polarization. For example, the angular distribution of the 
$b$ quarks in the top rest frame has the form
\beq
\frac{d\sigma}{d\cos\hat{\theta}_b}\,\propto\,
\left( \frac{m_t^2}{m_W^2} + 2\right)\left(E_\chi + \sin2\teff m_\chi\right)
\,+\, \left( \frac{m_t^2}{m_W^2} - 2\right)\,p_\chi\,\cos 2\teff\,\,
\cos\hat{\theta}_b\,,
\eeq{theta_b}
where $\hat{\theta}_b$ is the angle between the momenta of the $b$ quark and 
the neutralino coming from the same stop decay as the top, and $E_\chi$ and 
$p_\chi$ are the energy and momentum of this neutralino. (See Appendix A.) 
In the case of hadronic W decay, the top rest frame can be reconstructed. The 
neutralino direction is unknown, 
since there is at least two missing energy particles in each event. However, 
since stops are heavy, in the reaction~\leqn{direct} they are produced close 
to threshold, so that the stops in the direct sample tend to not 
have large velocities in the lab frame. If the stop is at rest in the lab 
frame, the top and neutralino from the same stop decay are back-to-back in this
frame. Then, the angle $\hat{\theta}_b$ is the same as $\theta_b$, the 
angle between the $b$ quark momentum in the top rest frame and the direction 
of the boost from this frame to the lab frame. The boost  
direction is simply the direction of the top momentum in the lab frame. The
angle $\theta_b$ is experimentally measurable. As we show below, the 
distribution of 
events in this angle provides a useful polarization analyzer, even though the
effect is washed out somewhat by the motion of the stops in the lab frame.

Another observable, useful for semileptonic top decays, is the angular
distribution of the charged lepton. In the top rest frame it has the form
\beq
\frac{d\sigma}{d\cos\hat{\theta}_l}\,\propto\,
E_\chi + \sin2\teff m_\chi+p_\chi\,\cos 2\teff\,\,
\cos\hat{\theta}_l\,.
\eeq{theta_l}
In semileptonic top decays, the missing neutrino does not allow for a
reconstruction of the top rest frame. (Since there are other unobserved particles in the event, the two neutralinos, the neutrino energy and 
momentum cannot be reconstructed by imposing the W and top mass constraints.)
However, one can define the ``approximate'' top rest 
frame in which ${\bf p}_b+{\bf p}_l = 0$, and the angular distribution of 
charged leptons in that frame can be used as a polarization analyzer, as will 
be shown below. Once again, we will use $\theta_l$, the angle 
between the lepton momentum in the (approximate) top rest frame 
and the direction of the boost from this frame to the lab frame,
as a stand-in for $\hat{\theta}_l$. This preserves most of the polarization 
asymmetry due to the low lab-frame velocities of directly produced stops.

Note that $t$ and $\bar{t}$ decays can be
combined in the polarization analysis, since the polarization effects 
are the same for the two samples as a consequence of CP conservation. 

\section{Observability of Top Polarization at the LHC: a Benchmark Study}
\label{sec:BP}

In this section, we demonstrate that top polarization 
in the stop pair-production sample can be observed and measured at the LHC.
As with most effects in the MSSM, observability of top polarization, and the
details of the analysis required to extract it from the data, depend on the 
choice of the MSSM parameters. We
will perform a detailed analysis at a specific benchmark point in the
MSSM parameter space, compatible with the golden region hypothesis of Ref.~\cite{GR}.

\subsection{Benchmark Point and Signature}

\begin{figure}
\begin{center}
\includegraphics[width=8cm]{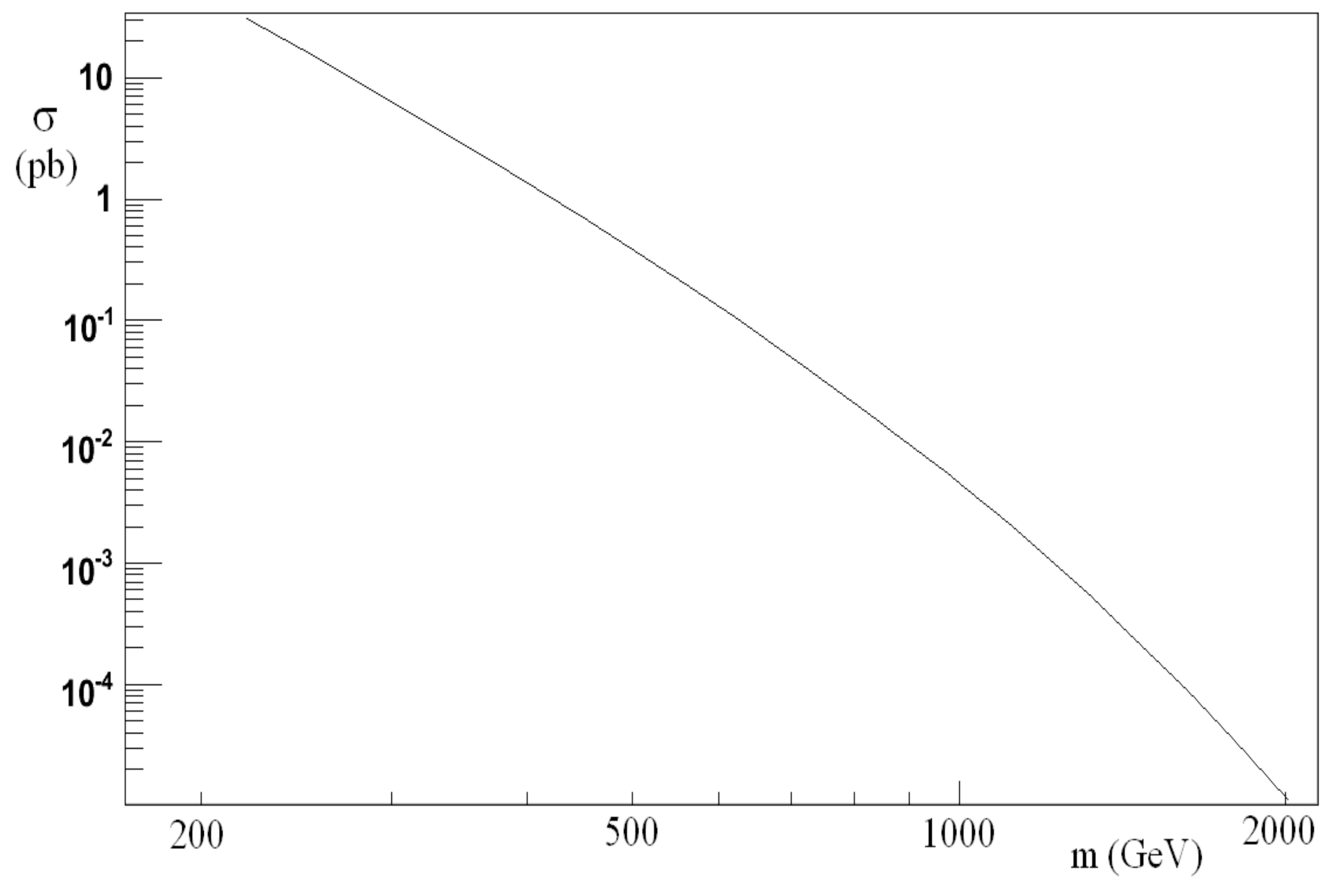}
\vskip2mm
\caption{Leading-order cross section for $pp\to\tilde{t}\tilde{t}^*$ at the LHC. We set the factorization and renormalization scales to $\mu = m_{\tilde{t}}$ and used the CTEQ6l1 parton distribution function set~\cite{cteq}. }
\label{fig:xsec}
\end{center}
\end{figure}

We choose the following benchmark point (BP) in the MSSM parameter space
(all parameters are specified at the weak scale):
\beqa
M_1&=&100~{\rm GeV},~~M_2\,=\,M_3\,=\,1~{\rm TeV}\,;~~~
\mu = 400~{\rm GeV},~~\tan \beta \,=\, 10;\CR
m(\tilde{Q}_{1,2}) &=& m(\tilde{u}^c_{1,2})\,=\,m(\tilde{d}^c_{1,2,3})\,=\,
m(\tilde{L}_{1,2,3})\,=\,m(\tilde{e}^c_{1,2,3})\,=\,1~{\rm TeV}\,.
\eeqa{MSSMpars}
All phases, flavor-violating couplings and masses, and all A-terms
except for $A_t$, are set to zero. The three stop sector lagrangian patameters,
$m(\tilde{Q}_3)$, $m(\tilde{u}^c_3)$, and $A_t$, determine the two
stop eigenmasses, $M(\tilde{t}_1)$ and $M(\tilde{t}_2)$, and the mixing
angle $\theta_t$. In our analysis, we dial the lagrangian parameters to
keep the stop masses fixed at
\beq
M(\tilde{t}_1)\,=\,340~{\rm GeV},~~~M(\tilde{t}_2)\,=\,800~{\rm GeV}\,,
\eeq{stopmass} 
while $\theta_t$ is varied to illustrate the analyzing power of the top 
polarization measurement. Note that the chosen stop eigenmasses, $\tan\beta$ 
and $\mu$ are compatible with the MSSM Golden Region, for values of $\theta_t$ 
sufficiently close to $\pi/4$ (maximal mixing). 

\begin{figure}
\begin{center}
\includegraphics[width=8cm]{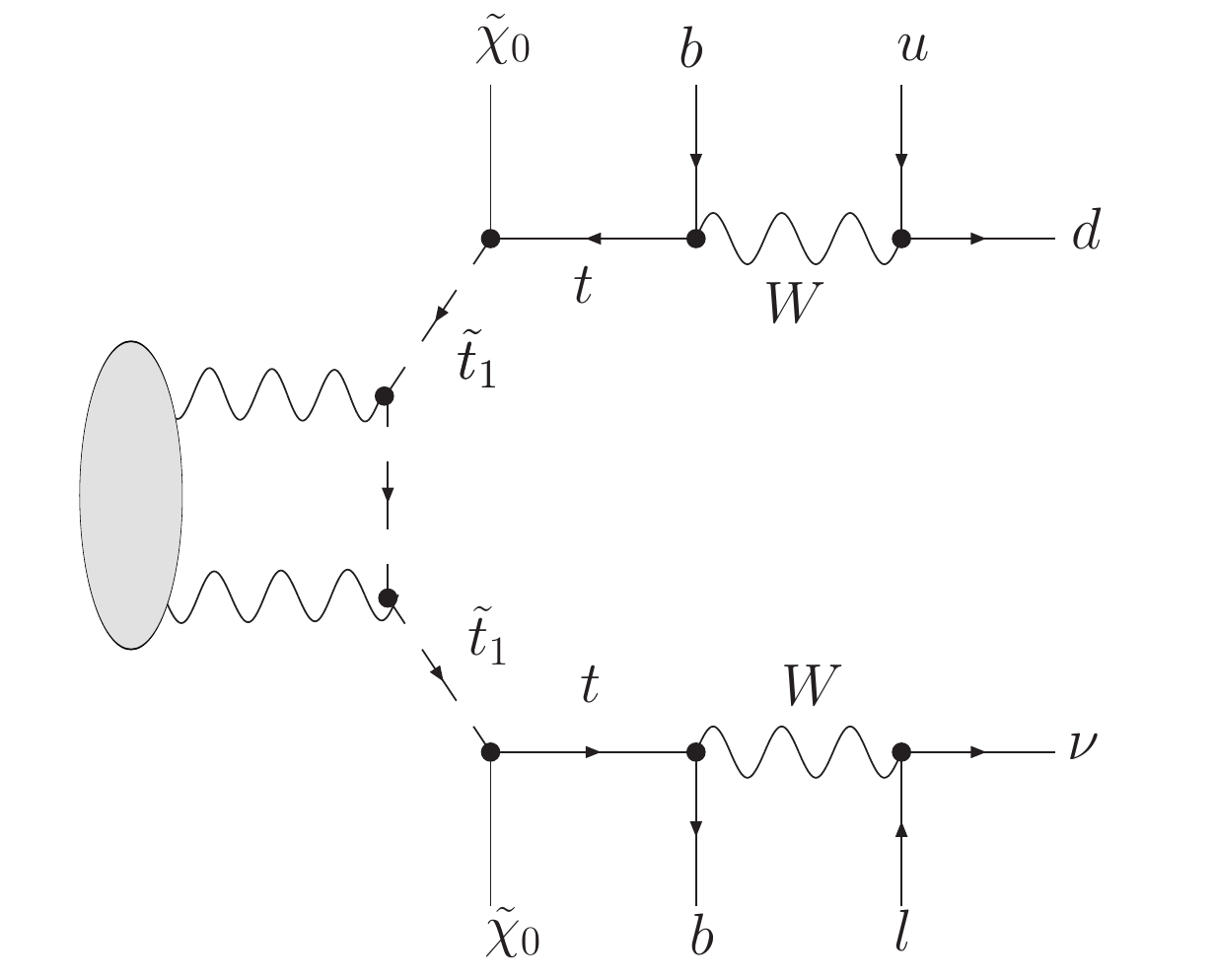}
\vskip2mm
\caption{One of the diagrams contributing to the direct stop production in proton-proton collision, followed by the decay chain leading to the final state~\leqn{signal}.}
\label{fig:feyn}
\end{center}
\end{figure}

The cross section of direct stop-antistop pair production at the LHC, as a function of the stop mass, is shown in Fig.~\ref{fig:xsec}. At the BP, the cross sections 
are\footnote{All signal cross sections cited in this paper are evaluated at leading
order, using CTEQ6l1 pdf set and setting the factorization and renormalization 
scale to the mass of the produced stop state. For the the $t\bar{t}$ background channels we set the factorization and renormalization 
scale to the top pole mass and for the $W+$jets events we use 
$\mu^2 = m_W^2+ \sum_{\rm jets} (m_{tr}^2)$, where $m_{tr}^2 = m^2+p_T^2$, summed over heavy quarks and light jets.
We use the {\tt MadGRAPH/MadEVENT}
software package~\cite{MG} to compute cross sections and generate events, except for the $W+$jets background channels, where we use {\tt ALPGEN}~\cite{AG}.} 
\beqa
\sigma(pp\to \tilde{t}_1\tilde{t}_1^*) &=& 3.23~{\rm pb}\,,\CR
\sigma(pp\to \tilde{t}_2\tilde{t}_2^*) &=& 33~{\rm fb}\,.
\eeqa{xsec}
The direct sample is completely dominated by $\tilde{t}_1$ pairs, and
we will ignore the small $\tilde{t}_2$ contamination in our analysis. 
Moreover, all indirect stop production processes, such as gluino decays, 
are strongly suppressed at our BP; we will ignore those processes.
The produced $\tilde{t}_1$ decays, with essentially 100\% probability, via
$\tilde{t}_1 \to t \chi_1^0$; all other two-body decays are kinematically 
forbidden. Thus, each event contains a top, an anti-top, and two neutralinos
in the final state. Both hadronic and semileptonic top decays provide 
observables which can be used to infer top polarization. We will focus on the
two angular distributions discussed in Section~\ref{sec:idea}, $\theta_b$ in 
hadronic top decays and $\theta_l$ for semileptonic tops. A priori, it is
not clear which observable would provide better sensitivity in a realistic 
experiment: While the hadronic channel suffers from the finite resolution of 
the jet energy measurement and combinatoric backgrounds, the leptonic 
channel is limited by the inability to reconstruct the top rest frame in 
each event. We will explore this question by analyzing events where one of
the tops decays hadronically and the other one semileptonically\footnote{For a recent study of the LHC reach for stops and spin-1/2 top partners in this channel, see Ref.~\cite{Han}.}, allowing us 
to study $\theta_b$ and $\theta_l$ distributions in the same sample. Thus, 
our final state is    
\beq
t(3j) + \ell + j + \met\,,
\eeq{signal}
where $\ell=e, \mu$. An example of a Feynman diagram contributing to this final state is shown in Fig.~\ref{fig:feyn}. Other final states (two hadronic or two semileptonic 
tops) can also be considered, and we expect that some of the techniques 
discussed here will be applicable to those channels as well. However a detailed
analysis is outside the scope of this paper. 

\subsection{SM Backgrounds and Cuts}

Before presenting the polarization analysis, let us 
discuss the SM processes contributing to this final state, and propose a set
of cuts that strongly suppress their contributions without materially affecting
the polarization asymmetries in the signal.

Several SM processes contribute to the final state~\leqn{signal}. The most 
important processes are listed in Table~\ref{tab:SBg}. For concreteness,
we will give cross sections with $\ell = \mu^+$.\footnote{For the signal and 
the $t\bar{t}$ background channels, the cross sections   
with $\ell = e^-, e^+, \mu^-, \mu^+$ are identical. For the $W+$jets 
background channel, the cross sections with a negatively charged lepton are 
suppressed by about a factor of 2 compared to the ones shown in 
Table~\ref{tab:SBg}.} 
The largest contribution by far comes from the $t\bar{t}$ channel, where one 
of the tops decays hadronically and the other one semileptonically. The 
semileptonic decays may produce $\mu^-$ directly or in a $\tau^-$ decay; we
list the two contributions separately, to emphasize that the cut efficiencies for the two contributions differ due to their different kinematics. Another
important background process is $4j+W$, where the invariant mass of three 
of the jets accidentally coincides with the top mass, and the $W$ decays 
leptonically. In contrast, another obvious background, $t\bar{t}Z$, where 
the $Z$ decays invisibly, has a very small cross section (only about 0.7 pb)
and thus we will not include it in the analysis.   

\begin{table}[t!]
		\renewcommand{\arraystretch}{1.2}
\begin{center}
\begin{tabular}{|l||r|r|r|r|r|} \hline
  & $\sigma_{\rm tot}$ & $\sigma_{\rm tot}\cdot {\rm Br}$ & 
$\sigma_{\rm tot}\cdot\,{\rm Br}\,\cdot\,{\rm Eff}$ 
& $N_{\rm sim}$ & Generator 
\rule{0ex}{2.2ex} \\ \hline \hline
$\tilde{t}_1\tilde{t}_1^*$ (BP) & 3.23    & 0.18 & 0.014 & 8270 & {\tt MG/ME} \\ \hline
$t\bar{t} (\mu^+)$              & 550     & 29.2 & $5.0 \cdot 10^{-3}$ & $1.2 \cdot 10^6$ & {\tt MG/ME}\\ 
$t\bar{t} (\tau^+ \to \mu^+)$   & 550     & 2.42 & $0.13 \cdot 10^{-3}$ & $2.2 \cdot 10^5$ & {\tt MG/ME}\\ 
$2j+2b+W^+$                     & 101     & 3.7 & $0.48 \cdot 10^{-3}$ & $3.3 \cdot 10^5$ & {\tt ALPGEN} \\ 
$4j+W^+$                        & 1330    & 132 & $0.90 \cdot 10^{-3}$ & $2.9 \cdot 10^5$ & {\tt ALPGEN} \\ \hline
Total BG                        & 2531    &  167   &    $6.5 \cdot 10^{-3}$  & $2.1 \cdot 10^6$   & \\ \hline
\end{tabular} \\[1ex]
\caption{Signal and Background cross sections (in pb), before and after cuts. 
Also listed are the total number of events simulated for our study (including the b-tag efficiencies), and the
software package used to generate the events.}
\label{tab:SBg}
\end{center}
	\renewcommand{\arraystretch}{1.}
\end{table}

\begin{figure}
\begin{center}
\includegraphics[width=6cm]{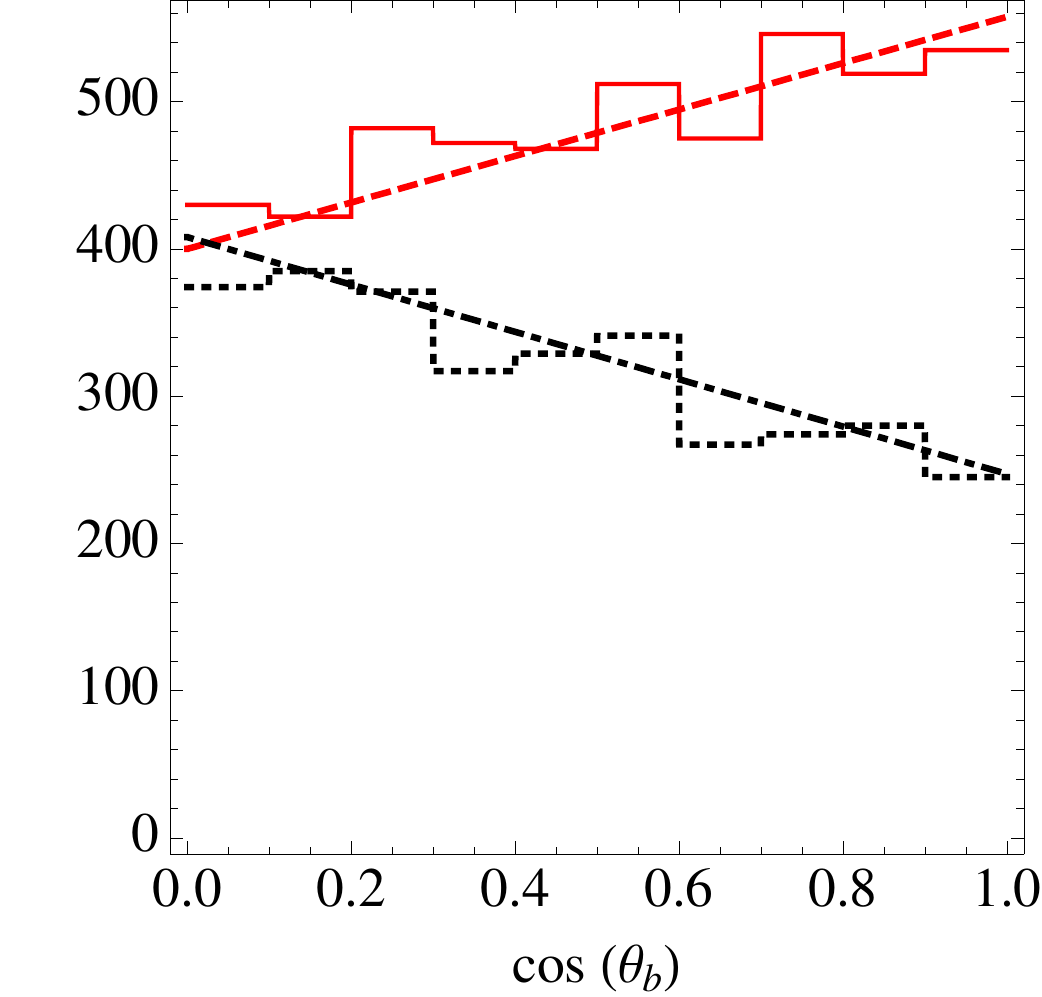}\hskip1cm
\includegraphics[width=6cm]{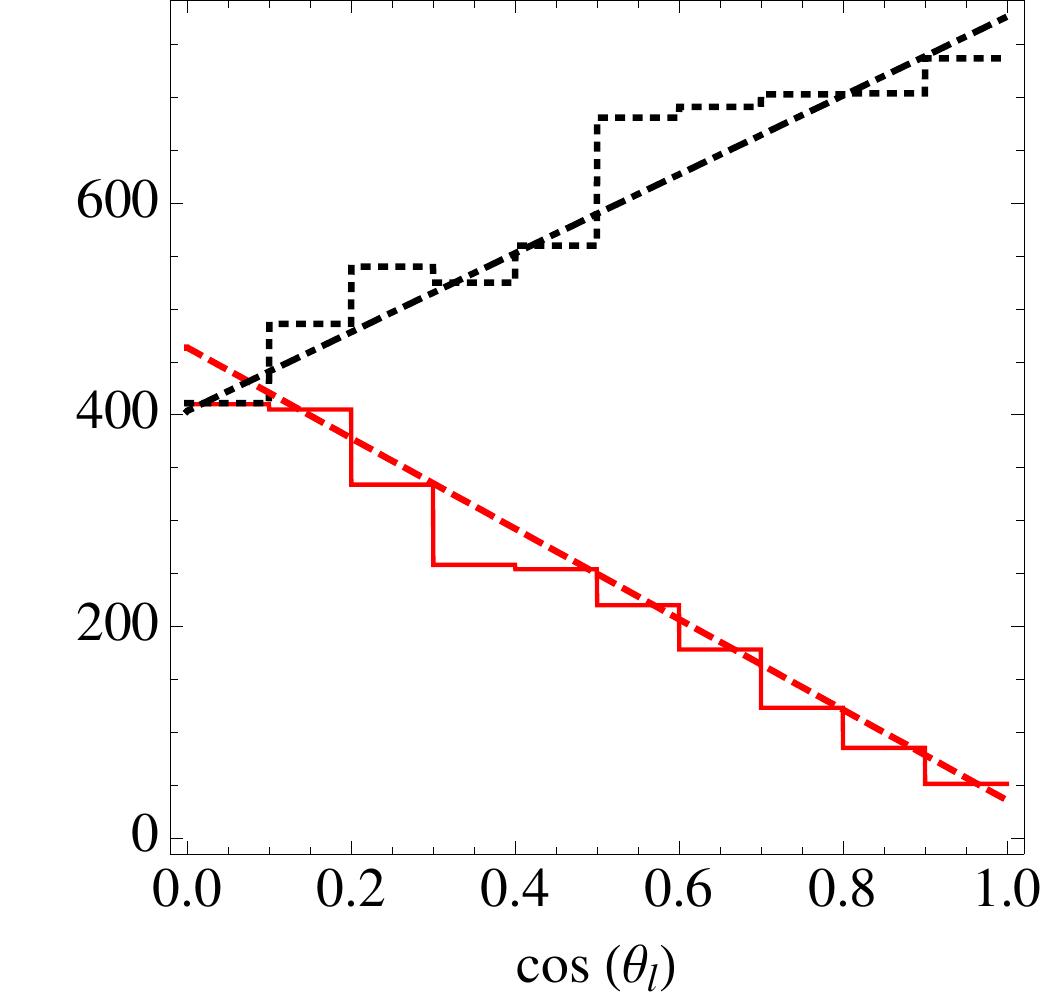}
\vskip2mm
\caption{Comparison of the analytic predictions and events generated by {\tt MadGraph/MadEvent} for the angular distribution of the $b$ (left) and
the charged lepton (right) in the top rest frame. Here $\hat{\theta}_b$ ($\hat{\theta}_l$) is the angle between the momenta of the $b$ quark (charged lepton) and the neutralino coming from the same stop decay as the top. The solid (red) histogram corresponds to the Monte Carlo events with $\cos 2\theta_{\rm eff} = -1$ and the dashed (red) line corresponds to the analytic prediction. The dashed and dash-dotted (black) lines correspond to the MC distribution and analytic prediction, respectively, for $\cos 2\theta_{\rm eff} = +1$.}
\label{fig:theo}
\end{center}
\end{figure}

We generated and analyzed Monte Carlo event samples for the signal and each SM 
background roughly corresponding to 10 fb$^{-1}$ of integrated luminosity at 
the LHC. The simulations were performed using {\tt MadGraph/MadEvent} (MG/ME)
software 
package~\cite{MG} except in the case of $W+$jets background, where we used 
the {\tt ALPGEN} package~\cite{AG}. The MG/ME simulations included integration
over the phase space of the fully decayed final state (8-body final state in
the case of signal, 6-body for the $t\bar{t}$ backgrounds). In the
matrix element calculations, small-width approximation was not used, but
only the Feynman diagrams corresponding to the desired decay chain, such as the one shown in Fig.~\ref{fig:feyn}, were 
included. 
All polarization effects in the decay chains of interest are 
fully taken into account in this approximation. 
To verify this, we compared the simulated distributions in the angles $\hat{\theta}_b$ and $\hat{\theta}_l$ for the signal to the analytic predictions in Eqs.~\leqn{theta_b},~\leqn{theta_l}. The results are in excellent agreement, as shown in Fig.~\ref{fig:theo}. Our analysis is restricted to 
the parton level; effects of hadronization, showering and intial state
radiation are not 
included. To roughly model the detector response to jets and electrons, we 
introduce a Gaussian smearing of their energies according to~\cite{TDRs}
\beq
\frac{\Delta E_j}{E_j} \,=\, \frac{50\%}{\sqrt{E_{\rm GeV}}}\,\oplus\,3\%\,,~~~
\frac{\Delta E_e}{E_e} \,=\, \frac{10\%}{\sqrt{E_{\rm GeV}}}\,\oplus\,0.7\%\,.
\eeq{smear}
We do not apply smearing to muon energies, since the effect 
is small. To model the acceptance of the detector, we apply the following
{\it acceptance cuts} at the generator level:

\begin{itemize}

\item $p_{T, j}>25$ GeV, $\eta\leq 4.0$ for each jet;  

\item $p_T^\ell \geq 20$ GeV and $\eta\leq 2.5$ for the charged lepton;

\item $\Delta R(j_i,j_k) \geq 0.4$ for each jet pair, and $\Delta R(j_i,\ell) 
\geq 0.4$ for the charged lepton and each jet ($i,k=1\ldots 4$).

\end{itemize}

\begin{figure}
\begin{center}
\includegraphics[width=6cm]{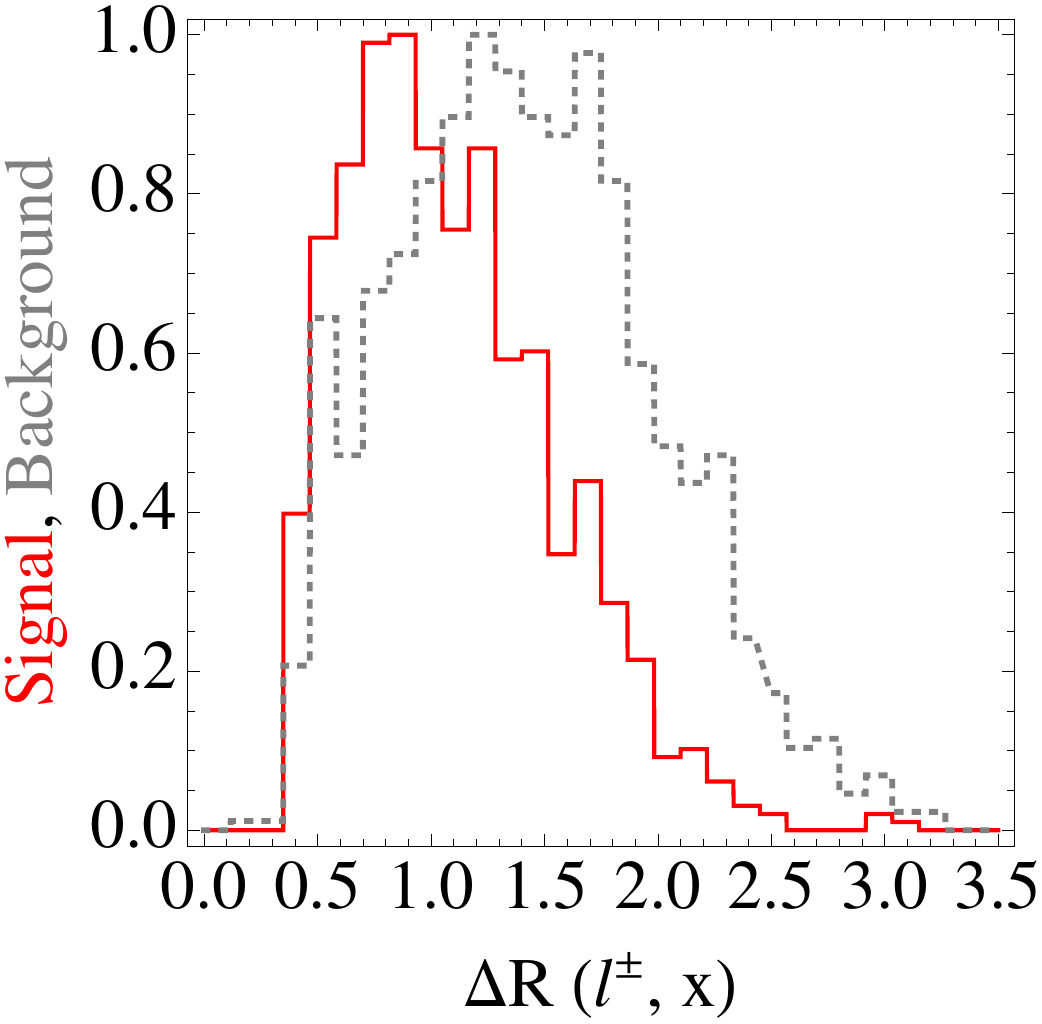}\hskip1cm
\includegraphics[width=6cm]{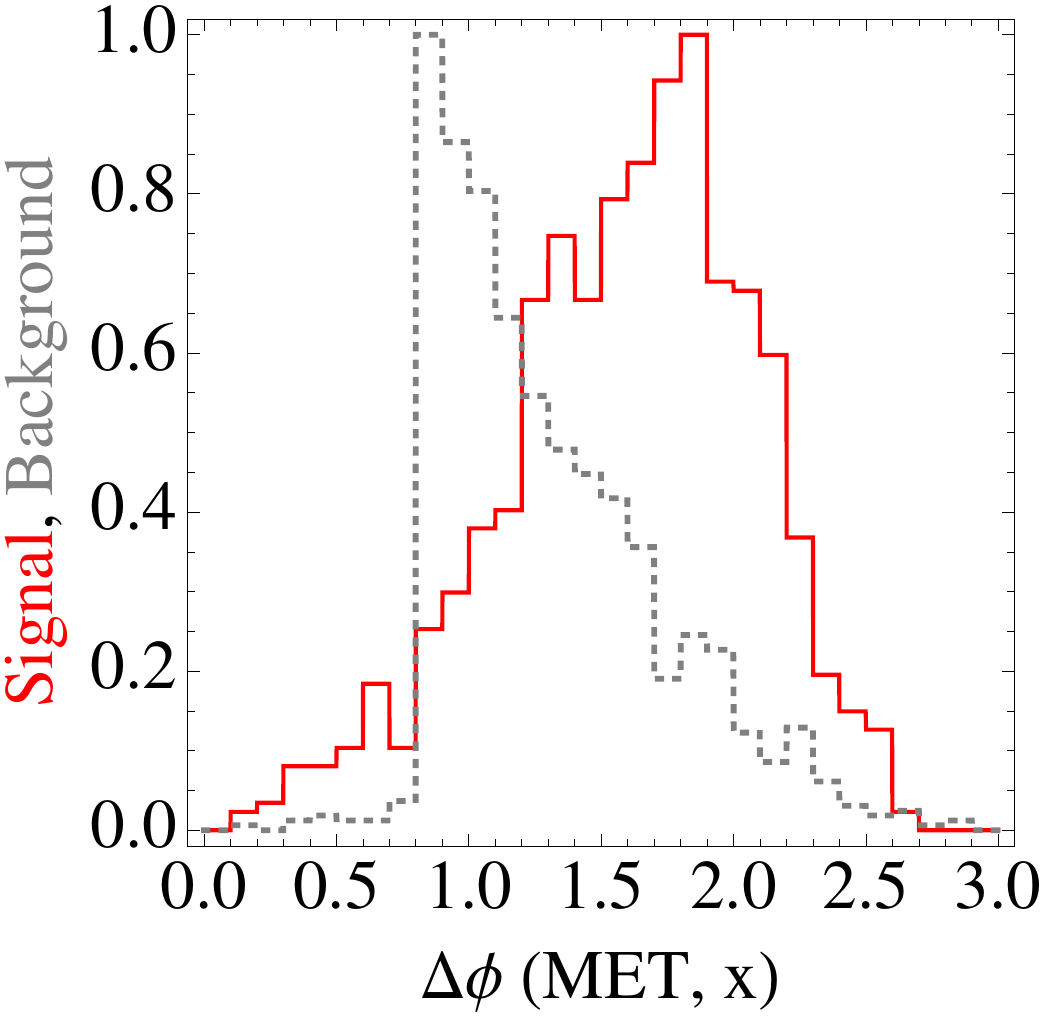}
\vskip2mm
\caption{Normalized distributions in $\Delta R(\ell^\pm, X)$ and $\Delta\phi(\met, X)$, in the signal and background samples satisfying all other acceptance and selection cuts. (See the text for the definition of the plotted observables.)  The solid (red) line corresponds to the signal and the dotted (gray)
line corresponds to the combined backgrounds.}
\label{fig:iso}
\end{center}
\end{figure}

The results of our analysis are summarized in Tables~\ref{tab:SBg}
and~\ref{tab:eff}. The first column of Table~\ref{tab:SBg} lists, for 
reference, the total cross section of each process. For the $W+$ jets 
processes, the listed cross sections are computed imposing a cut 
$p_{T, j}>20$ GeV on light (non-bottom) quarks and gluons. The second column of Table~\ref{tab:SBg}
lists the cross sections to produce the final state~\leqn{signal} in each 
channel, including the above acceptance cuts. At this point, the signal is
completely swamped by the $t\bar{t}$ background, with $S/B\sim 0.01$. To 
extract the signal, we impose the following {\it selection criteria}: 

\begin{itemize}

\item At least one of the jets must be $b$-tagged. We assume  energy-independent $b$-tag 
probabilities of 50\% for a true $b$-jet, 10\% for $c$-jet, and 
1\% for light quark and gluon jets.

\item Large missing transverse energy: $\met \geq 125$ GeV.

\item $\cos\phi(p_T^\ell,\met) \leq 0.7$, where $\phi(p_T^\ell,\met)$ is 
the opening angle between the transverse momentum of the charged lepton and 
the missing transverse momentum. This cut eliminates events where all missing 
energy comes from the leptonic decay of a highly boosted $W$.

\item Hadronic Top Reconstruction (HTR): $\min |m_{3j} - m_t^{\rm true}|\leq 
5$ GeV, where $m_{3j}$ is the invariant mass of three jets, and the minimum is 
over the four possible triads. In our study, the true top mass 
$m_t^{\rm true}$ is taken to coincide with the value used in the MC generator,
$172.5$ GeV. The jet triad for which the minimum is achieved is identified 
with the hadronic top. This cut suppresses $W+$jets backgrounds. 
 
\item Semileptonic Top Veto (STV): Assuming that all missing energy comes 
from a massless neutrino, we reconstruct the candidate neutrino four-momentum 
$\bar{p}^\nu$ using the transverse $\met$ measurement and the conditions 
$(\bar{p}^\nu)^2=0, (p^\ell+p^j+\bar{p}^\nu)^2=m_t^2$, where $p^j$ is the 
momentum of the jet which does {\it not} belong to the triad identified as the 
hadronic top. There are generically two solutions, $\bar{p}^\nu_{1,2}$. 
We then form $s^{\nu\ell}_i = (p^\ell+\bar{p}^\nu_i)^2$, and demand
$\min_{i} |\sqrt{s^{\nu\ell}_i} - m_W| \geq 40$ GeV. This cut suppresses 
the $t\bar{t}$ backgrounds, since the rejected kinematics corresponds to 
missing energy coming exclusively from semileptonic top decays.

\item Separation Cuts (SC): $\Delta R(\ell^\pm, X)\leq 1.5$ , where $\Delta R(\ell^\pm, X)$ is defined as the $\Delta R$
between the charged lepton and the jet closest to it in $\Delta R$; 
and $0.8 \leq \Delta\bar{\phi}(\met, X) \leq 1.3$, where $\Delta\bar{\phi}(\met, X)$ is defined as the azimuthal angle 
between the $\met$ vector and the object closest to it in $\phi$. The
signal and background distributions motivating these cuts are presented in Figure~\ref{fig:iso}. 

\end{itemize} 

\begin{table}[t!]
	\renewcommand{\arraystretch}{1.2}
\begin{center}
\begin{tabular}{|l||r|r|r|r|r||r||r|} \hline
  & $\met$ & $\cos\phi(p_T^\ell,\met)$ & HTR & STV & 
SC & b-tag  & 
Total  
\rule{0ex}{2.2ex} \\ \hline \hline
$\tilde{t}_1\tilde{t}_1^*$ (BP)  & 55    & 73   & 62   & 66   & 66     & 77   &  7.4  \\ \hline
$t\bar{t} (\mu^-)$               & 8.2   & 37  & 63   & 7.7  & 26  & 77   &  0.025   \\ 
$t\bar{t} (\tau^- \to \mu^-)$    & 17    & 7.4 & 61   & 5.2 &  7.1  & 77   &  $ 2.2\cdot 10^{-3}$   \\
$2j+2b+W^-$                      & 4.4   & 38  & 7.5  & 38 & 26    & 76   & $4.4\cdot 10^{-3}$ \\ 
$4j+W^-$                         & 3.0   & 38  & 8.2  & 37 & 24    & 5.1  & $4.2\cdot 10^{-4}$ \\ 
 \hline
\end{tabular} \\[1ex]
\caption{Selection cut efficiencies, in \%. (Each cut is applied to the sample satisfying the acceptance cuts and all selection cuts listed above it in the text.)}
\label{tab:eff}
\end{center}
\renewcommand{\arraystretch}{1.}
\end{table}

\noindent The efficiencies of these cuts for the signal and each of the
background samples are collected in Table~\ref{tab:eff}, and the cross sections after cuts for each process are listed in the third column of Table~\ref{tab:SBg}. With these cuts, we obtain
\beq
S/B = 2.5;~~~S/\sqrt{B} = 36~~~({\cal L}_{int} = 10~{\rm fb}^{-1}).
\eeq{SB}
Thus, we conclude that the stop signal is easily observable above the SM
backgrounds after the selection cuts are imposed. We can then use this 
signal-dominated event sample for the polarization analysis. 

Before proceeding, let us comment on the reliability of the Monte Carlo predictions for backgrounds, such as the $t\bar{t}$ simulation used in this study.  
Convincing interpretation of any excess over the SM background as contribution from new physics, in a situation where $S/B\sim 1$, would require precise understanding of the assumptions used in the background predictions. Since the LHC will produce a very large sample of $t\bar{t}$ pairs, the MC generators used to simulate this background can be precisely calibrated with data. This calibration can be done using samples which are not expected to suffer any SUSY contamination, such as the events with two reconstructed hadronic tops and small $\met$. It can then be applied to samples where SUSY can contribute, e.g. events with large $\met$ and a charged lepton which are our main focus here. It would be interesting to understand quantitatively the expected accuracy of such a validation procedure; such an analysis is however outside of the scope of this paper.

\subsection{Polarization Analysis}

In Section~\ref{sec:idea}, we identified two observables, the angles 
$\theta_b$ and $\theta_l$, which are potentially sensitive to the top 
polarization in the direct stop sample. The angle $\theta_b$ is the angle between the $b$ jet in the jet triad identified as the hadronic top,
and the direction of the hadronic top momentum, in the rest frame of the hadronic top. To define $\theta_l$, we first define the approximate rest 
frame (ARF) of the semileptonic top, by adding the three-momenta of the
charged lepton and the jet that does {\it not} belong to the reconstructed 
hadronic top. Then, $\theta_l$ is 
the angle, measured in the ARF, between the charged lepton and the direction
of the ARF's motion with respect to the lab frame.

To measure the angle $\theta_b$, we need to find which of the three jets in the hadronic top is the $b$ jet. One way to do it would be to simply demand that one of the jets be $b$-tagged. An alternative method, which only relies on kinematics, is to compute the invariant masses of the three jet pairs inside the hadronic top, and to find the pair whose invariant mass is closest to $m_W$. The jet that does not belong to that pair can then be identified as the $b$ jet. We found that in our sample the kinematic method has a higher efficiency than the simple $b$-tag, and thus we will use this technique to compute $\theta_b$ in each event.

\begin{figure}
\begin{center}
\includegraphics[width=5cm]{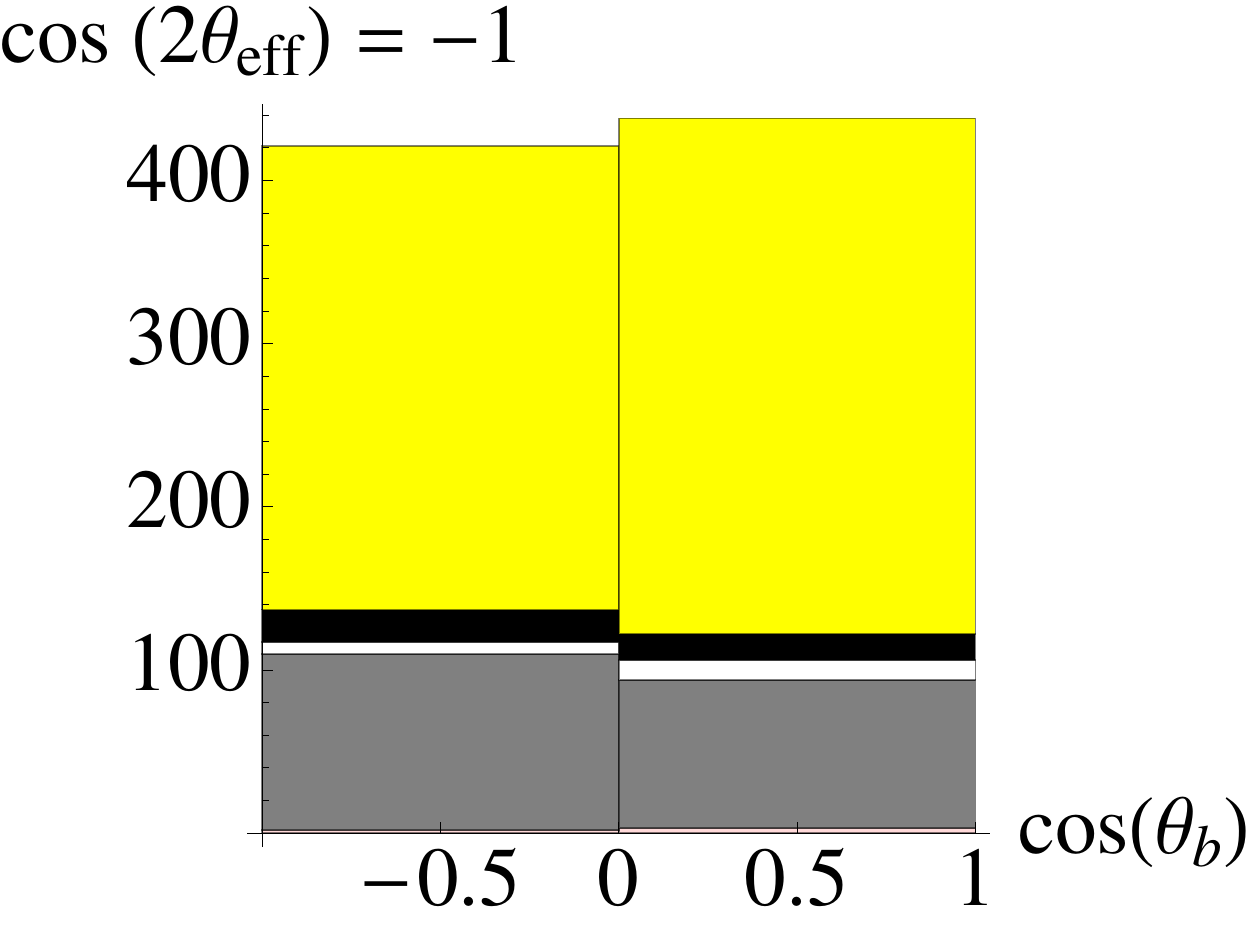}
\includegraphics[width=5cm]{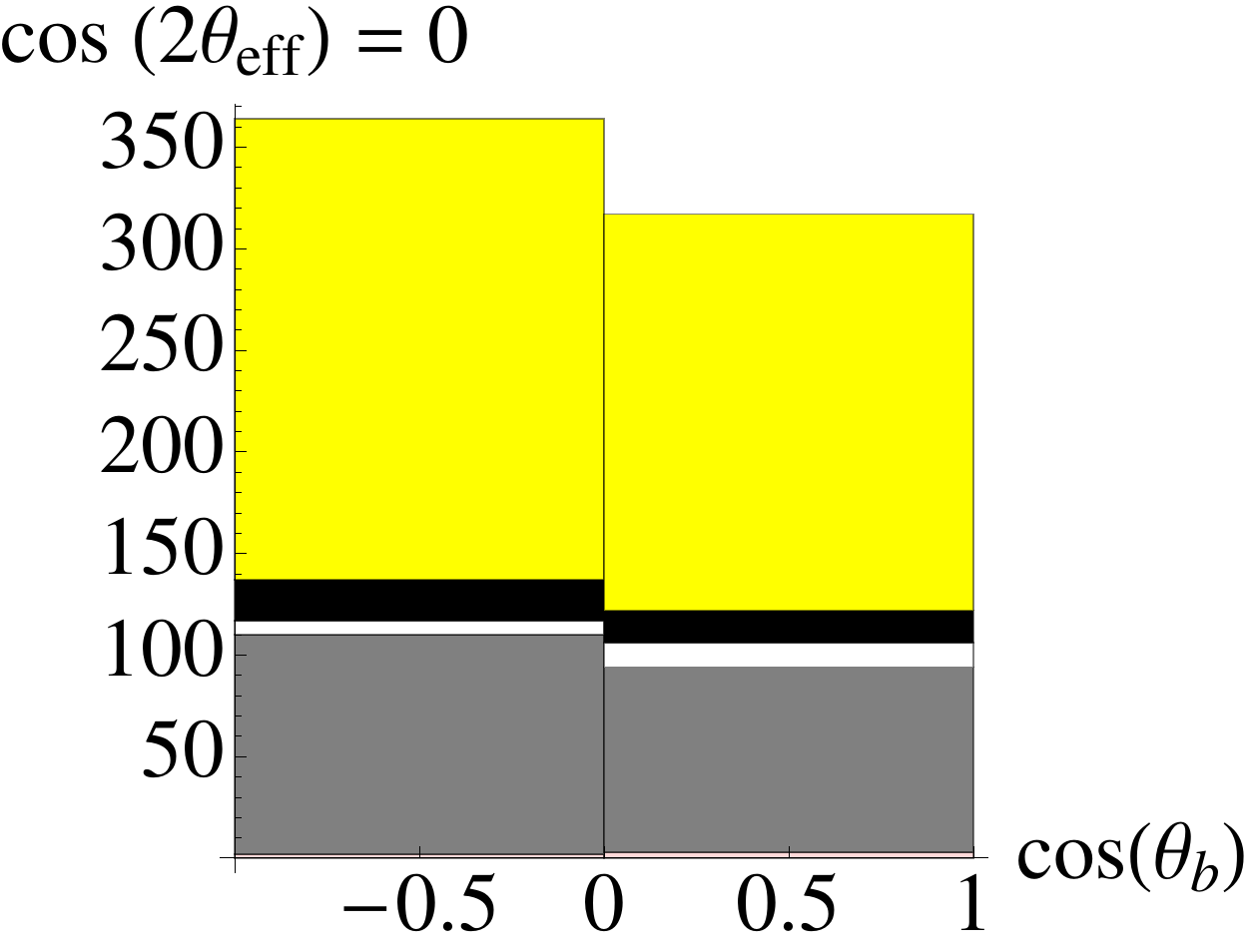}
\includegraphics[width=5cm]{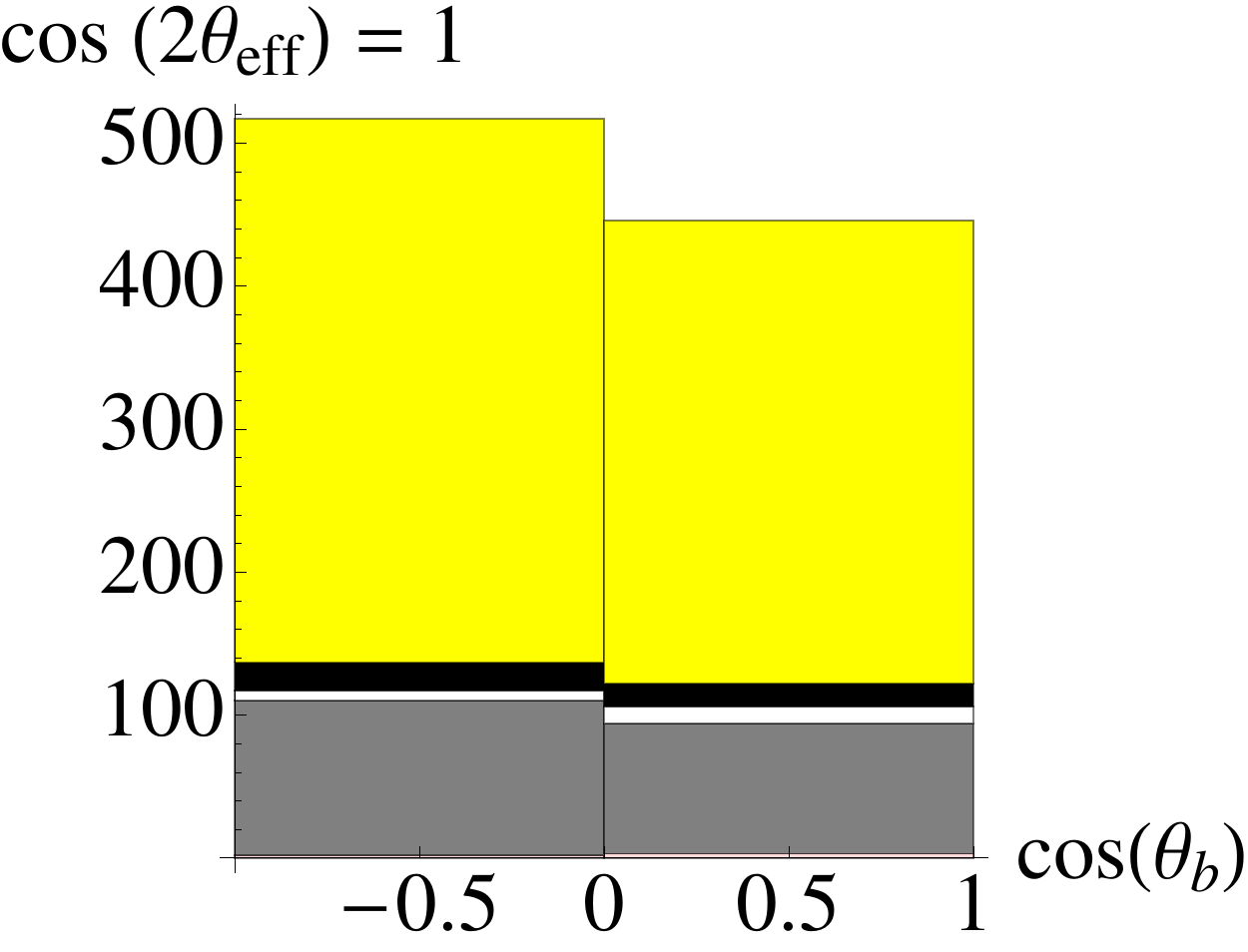}
\vskip2mm
\caption{Angular distributions of events in the angle $\theta_b$. The different
contributions correspond to (from top to bottom): signal (yellow), $4j+W^-$ (black), $2j+2b+W^-$ (white),
$t\bar{t} (\mu^-)$ (gray), $t\bar{t} (\tau^- \to \mu^-)$ (light red). The event numbers correspond to 10 fb$^{-1}$ integrated luminosity at the LHC.}
\label{fig:Ah}
\end{center}
\end{figure}

\begin{figure}
    \begin{center}
\includegraphics[width=5cm]{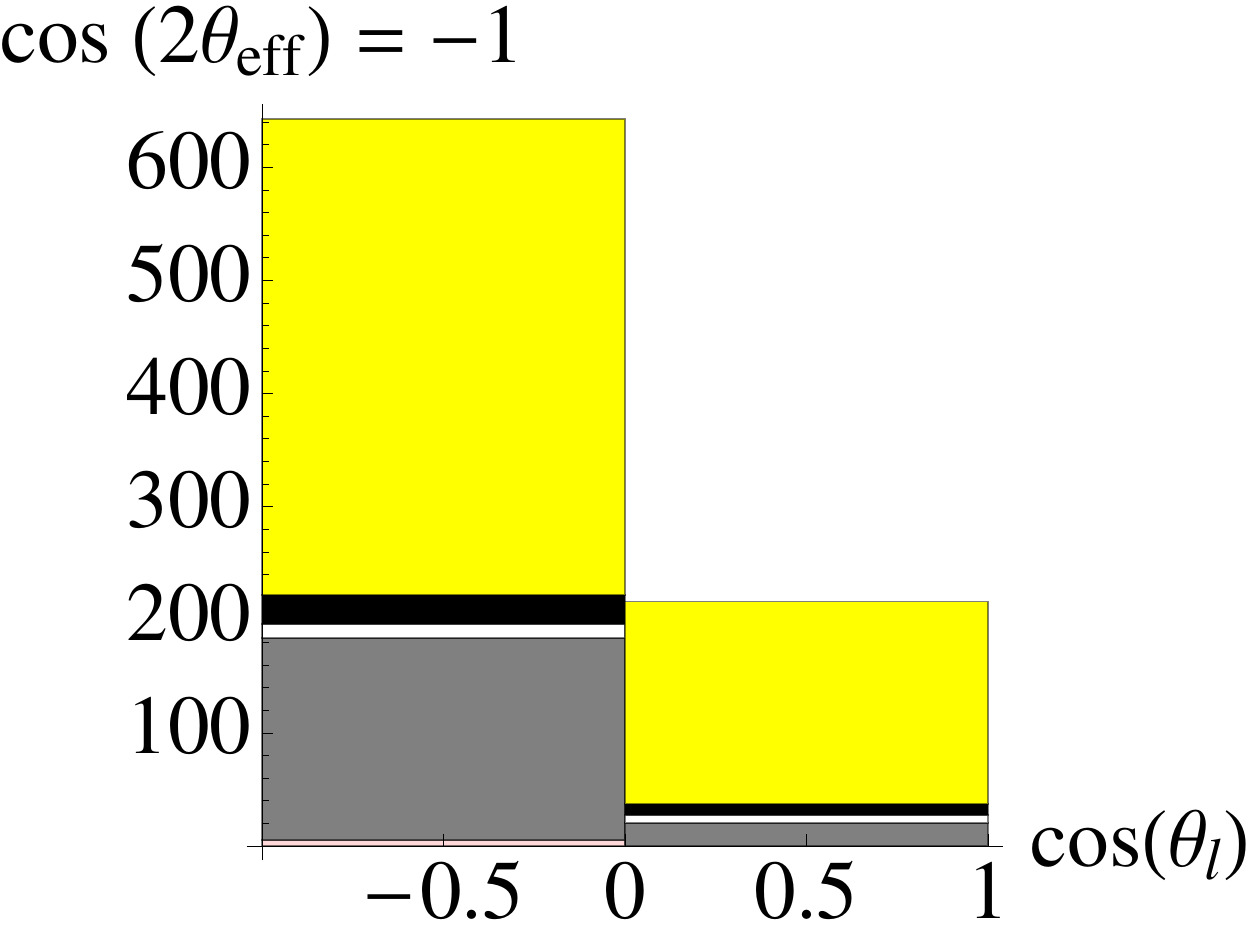}
\includegraphics[width=5cm]{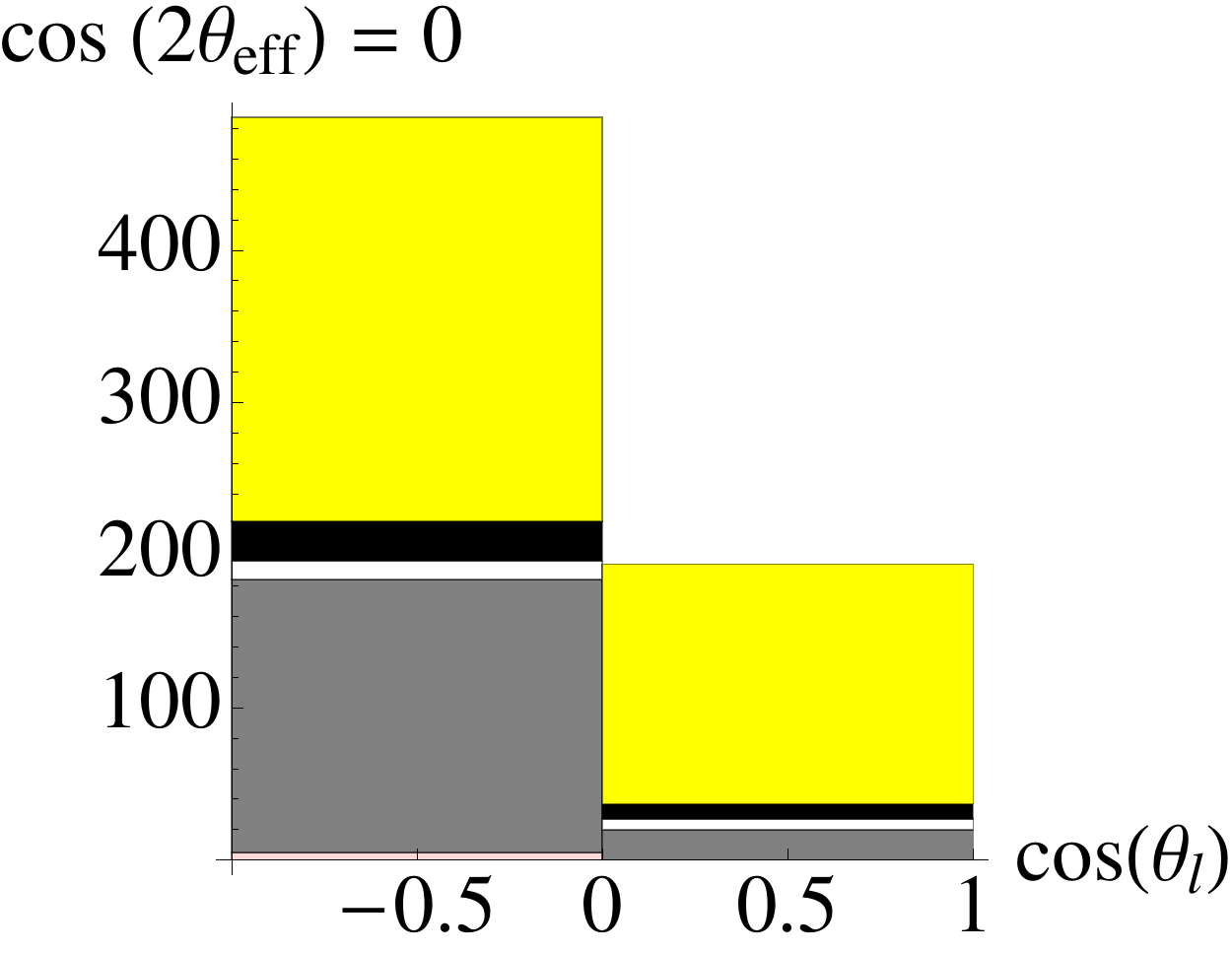}
\includegraphics[width=5cm]{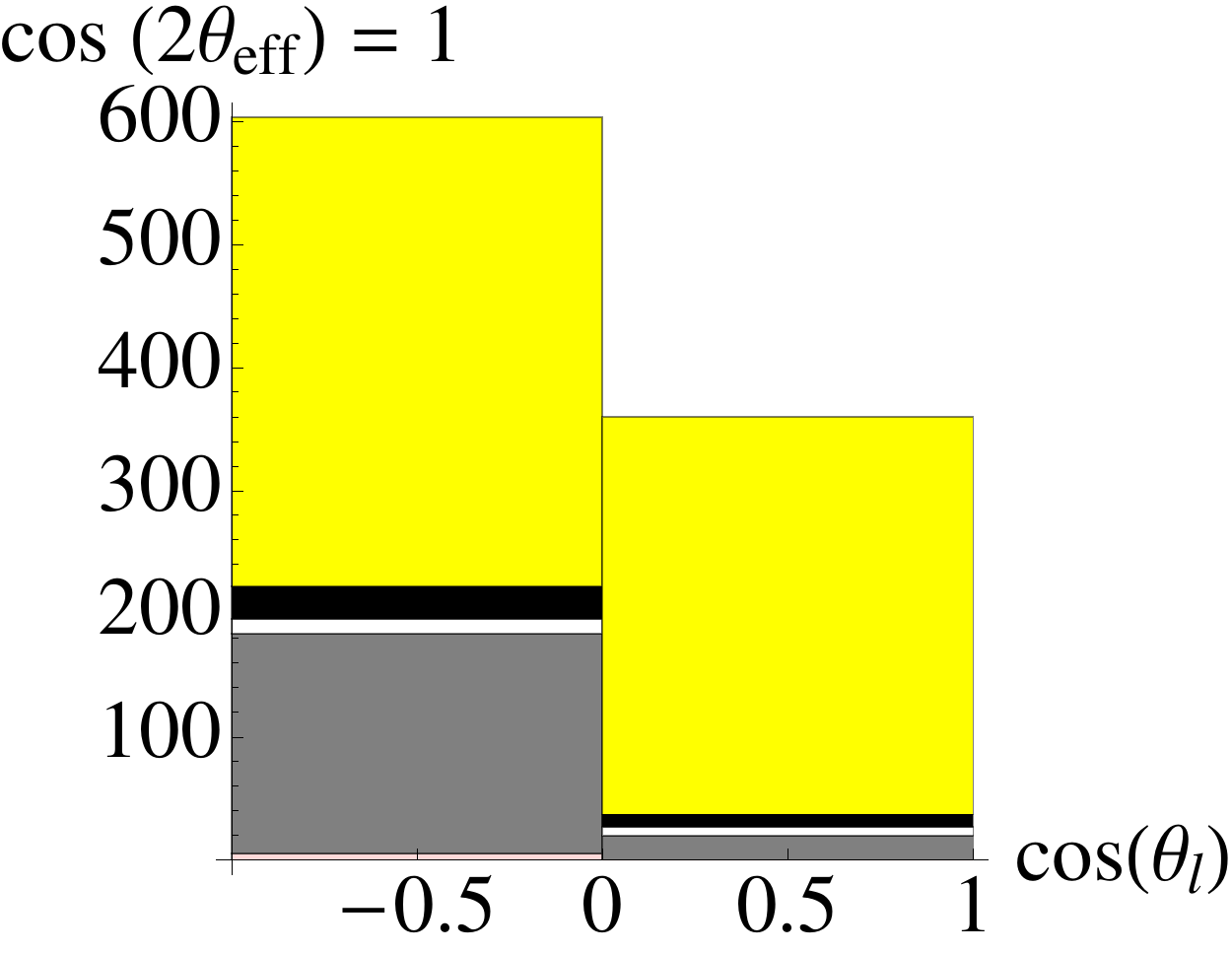}
\vskip2mm
\caption{Angular distributions of events in the angle $\theta_l$. The different
contributions correspond to (from top to bottom): signal (yellow), $4j+W^-$ (black), $2j+2b+W^-$ (white),
$t\bar{t} (\mu^-)$ (gray), $t\bar{t} (\tau^- \to \mu^-)$ (light red). The event numbers correspond to 10 fb$^{-1}$ integrated luminosity at the LHC.}
\label{fig:Al}
\end{center}
\end{figure}

In Figures~\ref{fig:Ah} 
and~\ref{fig:Al}, we show the distribution of events in our Monte Carlo 
sample (signal and background, passing the above acceptance and selection cuts) in 
$\cos\theta_b$ and $\cos\theta_l$. The event numbers correspond to 10 fb$^{-1}$ integrated luminosity at the LHC. Each figure shows the distributions for three values
of $\teff$: from left to right, $\teff=0, \pi/4, \pi/2$. The asymmetry due to 
top polarization is
clearly visible in the distribution. One should note that, even though we were careful to choose selection cuts that have as little effect on the hadronic asymmetry as possible, the cuts do have a slight preference for the events with $\cos\theta_b<0$, resulting in small asymmetries for the backgrounds and the signal at the parity-conserving point, $\teff=\pi/4$. Likewise, in the leptonic asymmetry, our use of the ARF instead of the true top rest frame leads to non-zero asymmetry in these cases. The important point, however, is that even though the measured asymmetry is not identical to the "primordial" asymmetry due to top polarization, there is still a simple one-to-one map between the two. To quantify the effect, we define two 
{\it forward-backward asymmetries (FBAs)}: The hadronic FBA is given by
\beq
A_{FB}^{\rm had} \,=\, \frac{\left(\int_0^1- \int_{-1}^0\right) d\cos\theta_b 
\,\frac{d\sigma}{d\cos\theta_b}}{\left(\int_0^1+ \int_{-1}^0\right) 
d\cos\theta_b \,\frac{d\sigma}{d\cos\theta_b}} \,, 
\eeq{Ahad_def} 
while the leptonic FBA is defined as
\beq
A_{FB}^{\rm lep} \,=\, \frac{\left(\int_0^1- \int_{-1}^0\right) d\cos\theta_l 
\,\frac{d\sigma}{d\cos\theta_l}}{\left(\int_0^1+ \int_{-1}^0\right) 
d\cos\theta_l \,\frac{d\sigma}{d\cos\theta_l}} \,.
\eeq{Alep_def} 
The statistical uncertainty of the FBA is given by 
\beq
\Delta A_{FB} \,=\,
2 \frac{\sqrt{N^+ N^-} } {(N^+ + N^-)^{\frac32} }
\eeq{Astat_def}
where $N^+=\int_0^1  d\cos\theta_i \,\frac{d\sigma}{d\cos\theta_i}$ and $N^-$ is the number of events in the opposite hemisphere.
The combined asymmetry is defined as
\beq
A_{FB}^{\rm combined} \,=\, (A_{FB}^{\rm had} - A_{FB}^{\rm lep}) -  \overline{ (A_{FB}^{\rm had} - A_{FB}^{\rm lep})}\, ,
\eeq{A_combined}
where the second term signifies the average over the $\cos 2\theta_{\rm eff}$
values.  
For our MC event sample, we obtain
\begin{center}
	\renewcommand{\arraystretch}{1.2}
\begin{tabular}{l||r|r|r|} 
                   & $\teff = 0$ & $\teff = \pi/4$ & $\teff = \pi/2$
\rule{0ex}{2.2ex} \\ \hline \hline
$A_{FB}^{\rm lep}$        & $ -0.49\pm 0.03   $       & $ -0.41\pm 0.03    $      & $-0.24\pm 0.03 $\\ 
$A_{FB}^{\rm had}$        & $ 0.022\pm 0.04  $      &$  -0.069\pm 0.04    $          & $ -0.074\pm 0.03$\\ 
$A_{FB}^{\rm combined}$   &$ 0.15\pm 0.05    $     & $-0.01\pm 0.05       $      & $-0.19\pm 0.04$\\ 
\end{tabular}
\renewcommand{\arraystretch}{1.}
\end{center}
where the errors are statistical only, corresponding to 10 fb$^{-1}$ worth of
data at the LHC. We conclude that, at the level of our analysis, polarization
effects in the stop direct sample are easily observable. 

\begin{figure}
\begin{center}
\includegraphics[width=5.3cm]{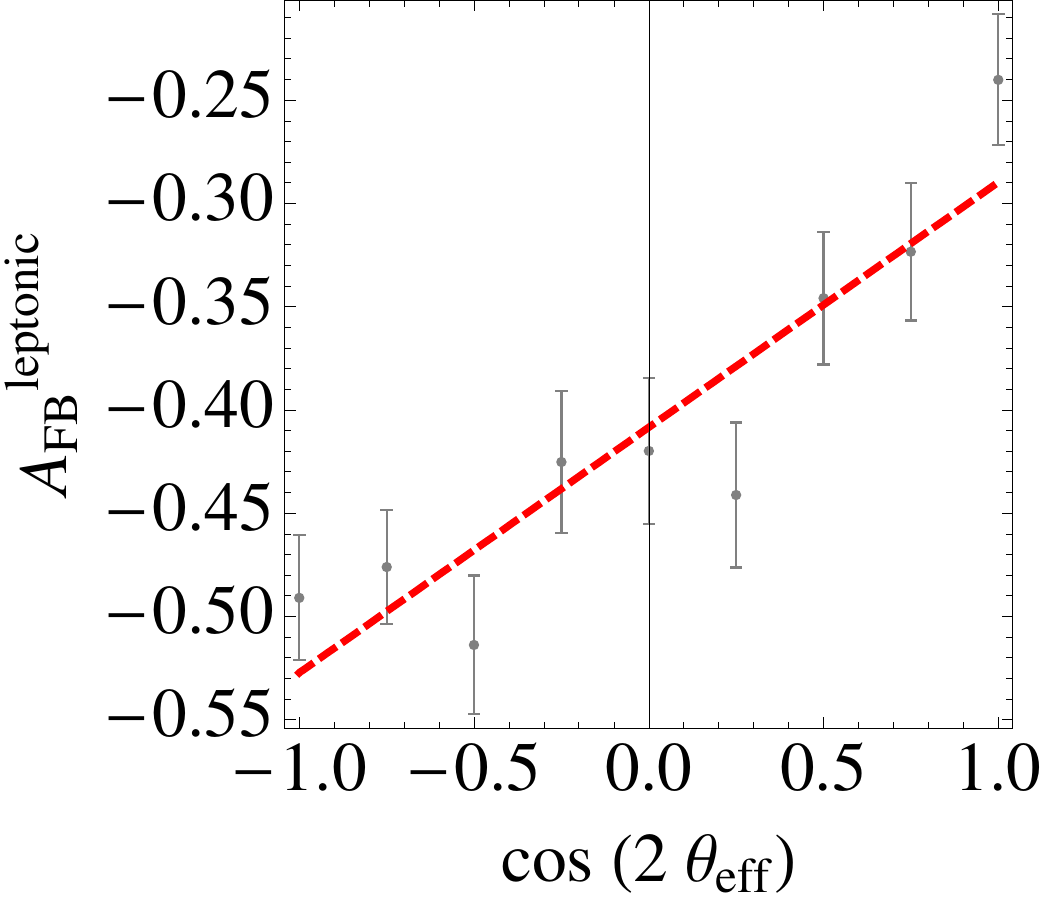}\hskip0.3cm
\includegraphics[width=5.3cm]{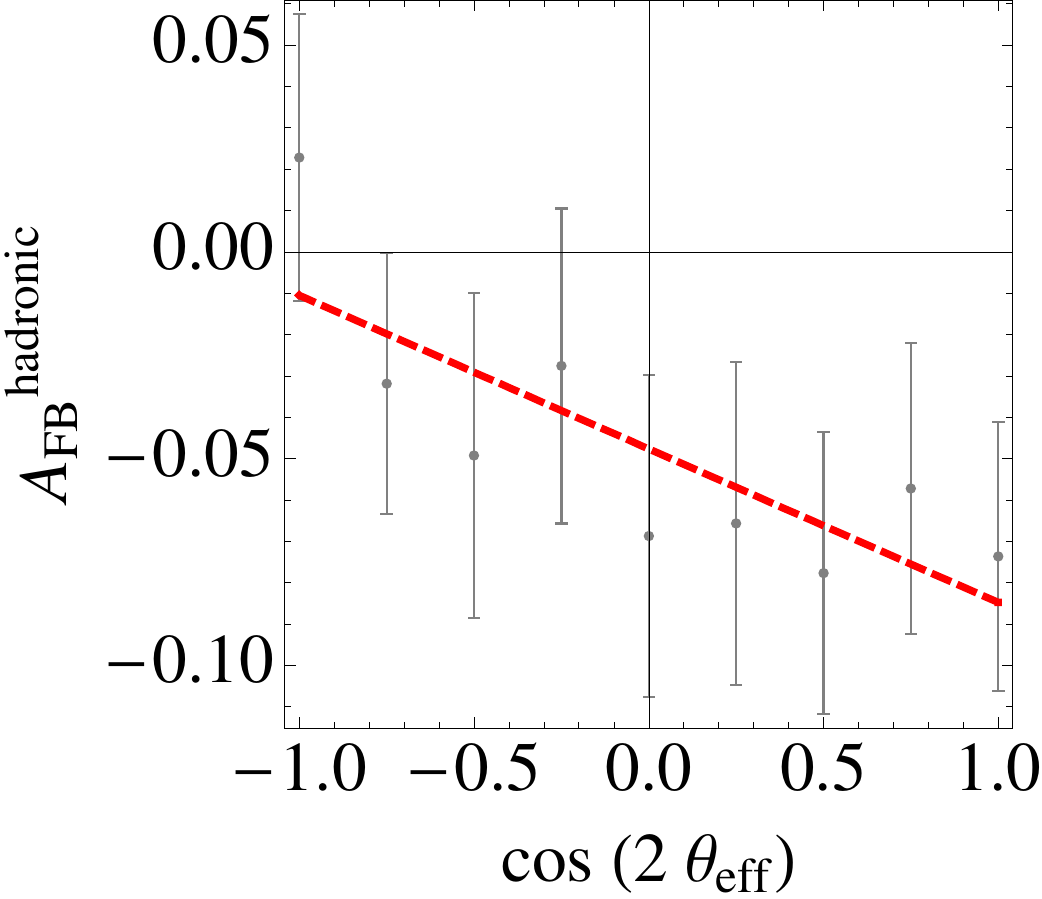}\hskip0.3cm
\includegraphics[width=5.3cm]{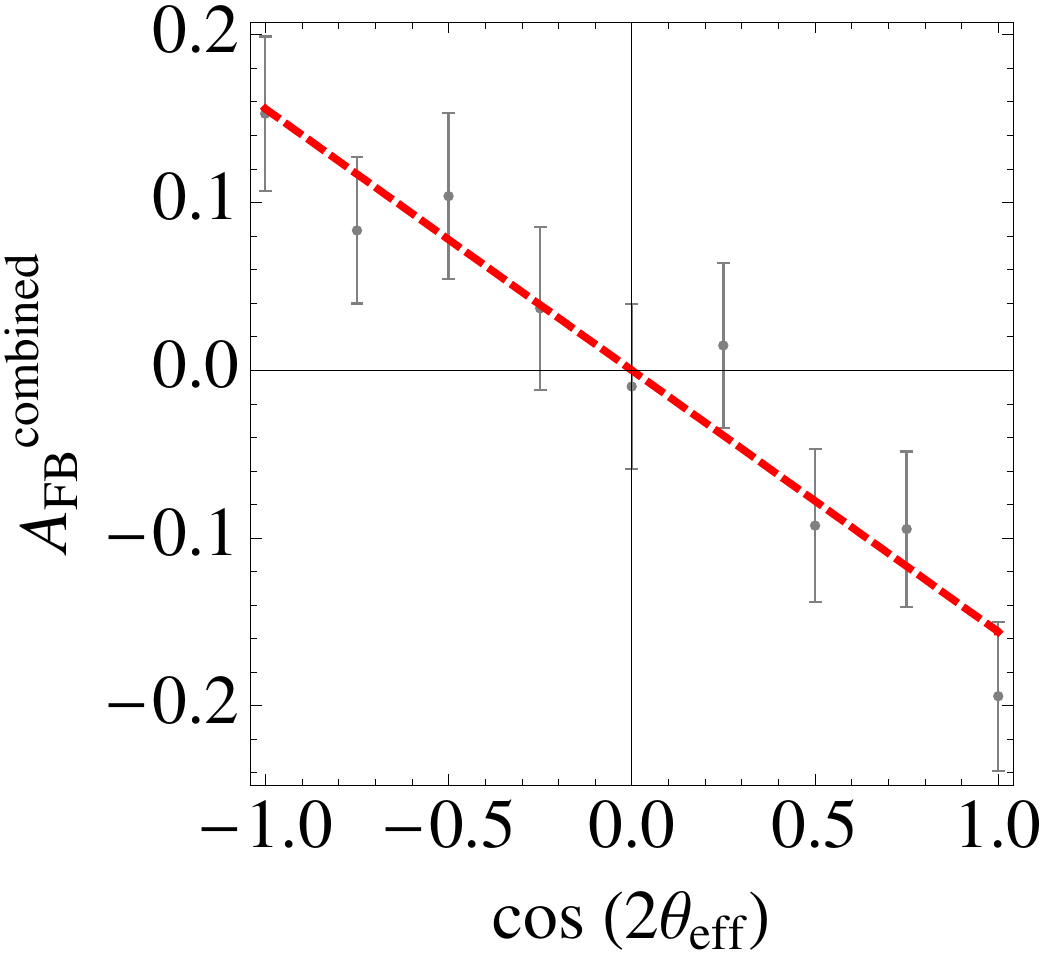}
\vskip2mm
\caption{Leptonic, hadronic, and combined forward-backward asymmetries, as a function of the angle $\teff$. The error bars indicate statistical errors for 
10 fb$^{-1}$ integrated luminosity.}
\label{fig:Asym}
\end{center}
\end{figure}

If the masses of $\tilde{t}_1$ and $\tilde{\chi}^0_1$ are measured 
independently, the measurement of the FBAs can be used to determine $\teff$. 
We studied the dependence of the hadronic and leptonic FBAs on $\teff$ by
generating MC samples of the signal for 10 values of $\teff$, keeping all
other MSSM parameters fixed at their BP values. The results are shown in 
Fig.~\ref{fig:Asym}. The error bars indicate statistical errors for 
10 fb$^{-1}$ integrated luminosity. As expected, we observe an approximately 
linear relationship between the asymmetries and $\cos2\teff$:
\beqa
A_{FB}^{\rm lep} &=& +0.12~\cos2\teff - 0.41\,,\CR
A_{FB}^{\rm had} &=& -0.037~\cos2\teff -0.047\,,\CR 
A_{FB}^{\rm combined} &=& -0.16~\cos2\teff\,.
\eeqa{slopes}
These formulas can be used to estimate the
precision of $\teff$ determination. Note, however, that the coefficients in these formulas depend on the stop and neutralino masses, and would need to be recalculated if these parameters differ from the BP values. The errors in the experimentally determined masses would need to be taken into account in the $\teff$ determination.

\begin{figure}
\begin{center}
	\includegraphics[width=5.3cm]{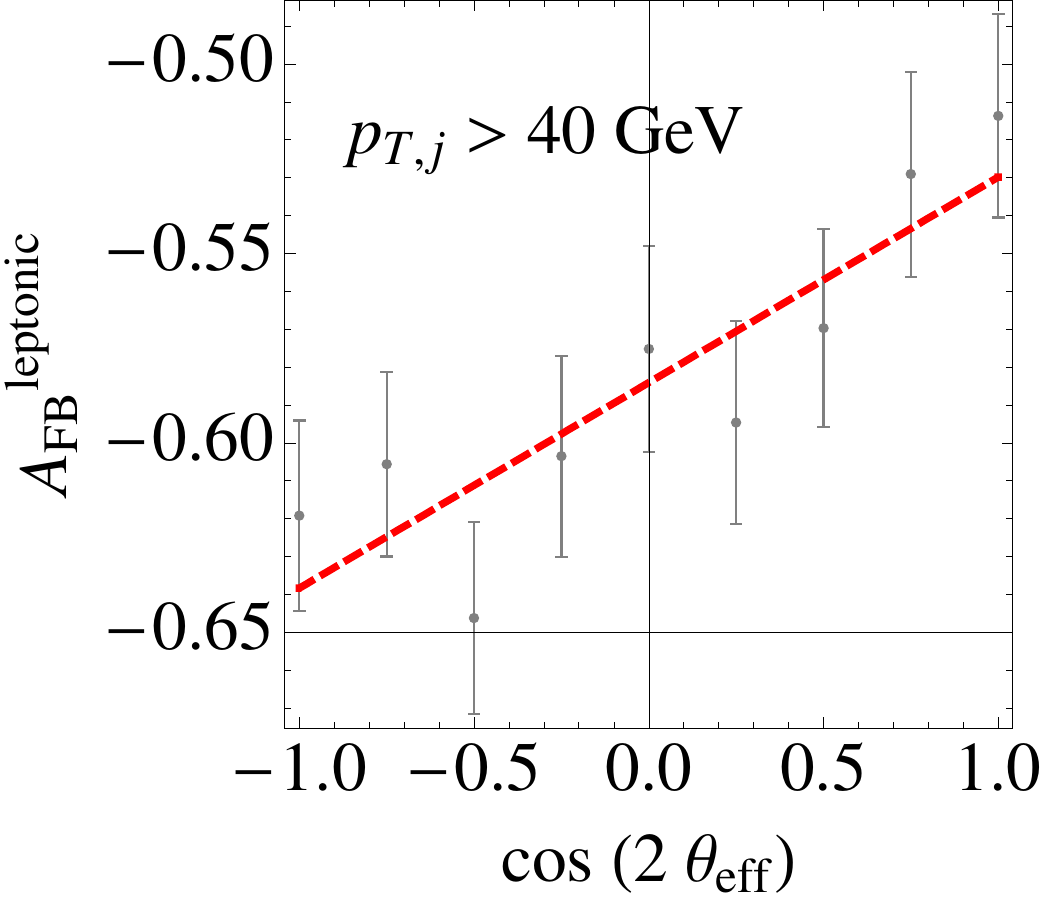}\hskip0.3cm
	\includegraphics[width=5.3cm]{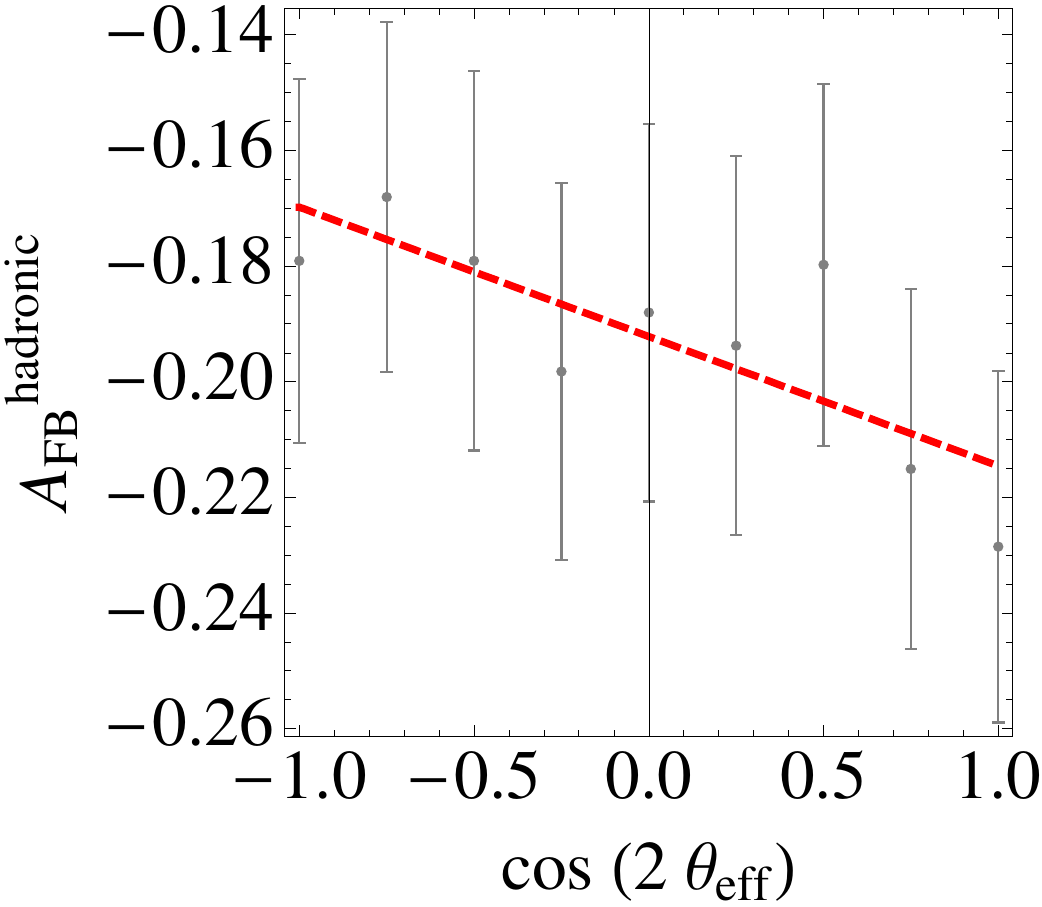}
\vskip2mm
\caption{Leptonic and hadronic forward-backward asymmetries for $p_{T,j}>40$ GeV, as a function of 
the angle $\teff$. The error bars indicate statistical errors for 
10 fb$^{-1}$ integrated luminosity.}
\label{fig:Asym40}
\end{center}
\end{figure}

Among the effects not included in our parton-level analysis, arguably the most important one is the absence of extra jets emitted as initial state radiation (ISR). Such jets would introduce a combinatoric background, making it more difficult to identify hadronic tops and potentially washing out the forward-backward asymmetries. The $p_T$ distribution of the ISR jets is sharply peaked at zero, since most of these jets are soft or collinear with the incoming beams. An additional lower cut on the jet $p_T$ (above the acceptance cut of $p_{T,j}\geq 20$ GeV assumed in the above analysis) may be necessary to suppress such jets. To estimate the sensitivity of our results to such a cut, we have repeated the analysis requiring $p_{T,j}\geq 40$ GeV, keeping all other cuts the same. The results are 
presented in Fig.~\ref{fig:Asym40}. It is clear that the hadronic asymmetry becomes essentially useless in this case, while the leptonic asymmetry is only marginally affected. The main effect of the additional $p_{T,j}$ cut on the leptonic asymmetry is in fact simply due to the reduced statistics, which can be overcome in time by increased integrated luminosity. Thus, we are optimistic that the leptonic asymmetry would survive as a useful observable even after the ISR and other showering, fragmentation and detector effects are included. Of course, a full detector-level analysis is needed to confirm this conclusion.

\section{From Top Polarization to Stop Mixing.}
\label{sec:teff}

\begin{figure}
\begin{center}
\includegraphics[width=10cm]{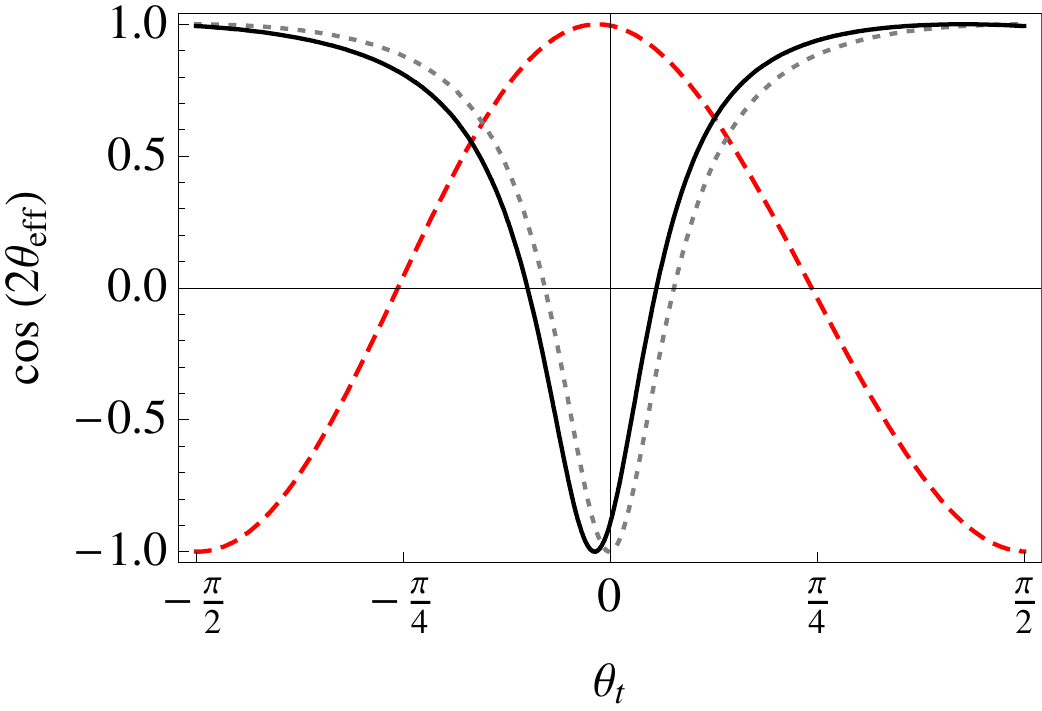}
\vskip2mm
\caption{$\cos 2\theta_{\rm eff}$ vs. the stop mixing angle $\theta$ for three different neutralino
mixing matrices. The solid (black) curve is our benchmark scenario, the dotted (gray) curve
corresponds to the case where the lightest neutralino is pure bino and the dashed (red) curve
corresponds to pure higgsino.}
\label{fig:thetaeff}
\end{center}
\end{figure}

If the effective angle $\teff$ is measured using the technique described in the previous sections, the actual stop mixing angle can be 
found by inverting Eq.~\leqn{thetaeff}:
\beq
\tan\theta_t = \frac{y_t N_{14} \cos\teff - \sqrt{2}\left( \frac{g}{2}N_{12} + 
\frac{g^\prime}{6} N_{11} \right) \sin\teff}{y_t N_{14} \sin\teff +
\frac{2\sqrt{2}}{3}g^\prime N_{11} \cos\teff}\,.
\eeq{invert_theta}
Note, however, that several issues arise in this conversion. First, the top polarization observables measure $\cos2\teff$, which introduces a two-fold degeneracy in the determination of $\theta_t$. Secondly, in some cases, the mapping from $\cos2\teff$ to $\theta_t$ introduces large errors even if $\cos2\teff$ is precisely known: for example, this can occur for pure-bino neutralino if $\cos2\teff$ is close to one, as seen in Fig.~\ref{thetaeff}. Finally, to determine $\theta_t$, one needs to know the composition of the lightest neutralino in terms of the gauge eigenstates. For example, in the cases of pure bino and pure higgsino, drastically different values of $\theta_t$ would be inferred from the same $\teff$ (see Fig.~\ref{thetaeff}). In the case when the neutralino is predominantly a bino, it is especially important to independently constrain the subdominant contributions, since the $\tilde{H} \tilde{t}t$ and $\tilde{W}^3 \tilde{t} t$ coupling constants are larger than the $\tilde{B}\tilde{t}t$ coupling, so that even a small higgsino or wino admixture can have a significant effect of $\teff$. Constraining the neutralino mixings at the level required to make the $\teff\to\theta_t$ conversion possible may in fact require substantially more integrated luminosity than the top polarization measurement itself. In spite of all these complications, this strategy seems to offer our best hope for the direct determination of the angle $\theta_t$ in the unconstrained MSSM framework, and is worth pursuing if SUSY is discovered.   

\section{Discussion and Conclusions}
\label{sec:conc}

In this paper, we argued that top and anti-top quarks produced in stop decays in the MSSM are generically expected to be polarized, and 
that observing and measuring this polarization can provide information about the stop mixing angle. We proposed observables sensitive to top polarization, and illustrated the potential of the LHC to observe this effect, by performing a detailed study for a single benchmark point in the MSSM parameter space. We conclude that for the chosen benchmark point, the polarization effect should be observable with
10 fb$^{-1}$ of data, and a useful measurement of the "effective" stop mixing angle $\teff$ can be obtained. We also briefly discussed how the measurement of $\teff$ can be used to constrain the actual stop mixing angle $\theta_t$, one of the crucial parameters in assessing the naturalness of the EWSB in the MSSM. 

The most important limitation of our analysis is that it is performed at the parton level, and only a very rough modeling of finite detector resolution by smearing jet and electron momenta was included. It is very important to perform a detector-level analysis including the effects of showering, hadronization and fragmentation, as well as detector effects. In particular, as we already remarked, ISR jets could contribute serious combinatoric backgrounds not included in this study, particularly for low-$p_T$ jets. These backgrounds can be suppressed by tightening the minimum $p_{T,j}$ cut, and we have verified that at least one of our observables (the lepton asymmetry) appears to be rather insensitive to such tightening (see Figs.~\ref{fig:Asym},~\ref{fig:Asym40}). Still, it is important to analyze these issues quantitatively with a detector-level analysis.  

\vskip0.5cm

\noindent{\large \bf Acknowledgments} 

We are grateful to Johannes Heinonen, Jay Hubisz, Mihoko Nojiri, Matthew Reece, Christian Spethmann and Gilad Perez for useful discussions. MP would like to thank the Institute for Physics
and Mathematics of the Universe (IPMU) in Tokyo, and MP and AW would like to thank the Kavli Institute for Theoretical Physics (KITP) in Santa Barbara, where parts of this work were completed. AW is
grateful to the Aspen Center of Physics where parts of this work were done.
 This research is supported by the NSF grant PHY-0355005.

{\it Note Added:} The main idea of this paper and some early results have been presented at the KITP Conference "Anticipating Physics at the LHC" in June 2008~\cite{MPtalk}. Related ideas have also recently appeared in Ref.~\cite{Shelton}.

\begin{appendix}

\section{Observables Sensitive to Top Polarization}

The top quark decays before it hadronizes, and the information about its
helicity is reflected in the distributions of the decay products. 
Consider the process $\t1\to\n1 t$, followed by the decay of the top quark 
$t\to W^+ b$. The matrix element for this process is proportional to
\beq
\bar{u}(p_\chi) \,\left( \cos\teff P_L + \sin\teff P_R \right) \,
\sum_{i=\pm} u_i(p_t) \bar{u}_i(p_t) \gamma^\mu P_L u(p_b) \epsilon_\mu 
(p_W)\,.
\eeq{M}
It is straightforward to calculate this matrix element in the coordinate 
system where the top quark is at rest, and the neutralino momentum is along 
the $+z$ axis. The non-zero matrix elements in the helicity basis are 
\beqa
{\cal M}(\chi_+, W_0) &=& \left( \cos\teff \sqrt{E_\chi - p_\chi} 
+ \sin\teff \sqrt{E_\chi + p_\chi} \right) \cdot \frac{m_t}{m_W}\,
\sin\frac{\theta}{2}\,,\CR
{\cal M}(\chi_+, W_\perp) &=& \left( \cos\teff \sqrt{E_\chi - p_\chi} 
+ \sin\teff \sqrt{E_\chi + p_\chi} \right) \cdot \sqrt{2}\,
\cos\frac{\theta}{2}\,,\CR
{\cal M}(\chi_-, W_0) &=& \left( \cos\teff \sqrt{E_\chi + p_\chi} 
+ \sin\teff \sqrt{E_\chi - p_\chi} \right) \cdot \frac{m_t}{m_W}\,
\cos\frac{\theta}{2}\,,\CR
{\cal M}(\chi_-, W_\perp) &=& \left( \cos\teff \sqrt{E_\chi + p_\chi} 
+ \sin\teff \sqrt{E_\chi - p_\chi} \right) \cdot \sqrt{2}\,
\sin\frac{\theta}{2}\,,\CR
\eeqa{Mhel}
where $E_\chi$ and $p_\chi$ is the energy and momentum of the neutralino, 
$\theta$ is the angle between the $b$ quark momentum and the $z$ axis. The
differential cross section is proportional to
\beq
\left( \frac{m_t^2}{m_W^2} + 2\right)\left(E_\chi + \sin2\teff m_\chi\right)
\,+\, \left( \frac{m_t^2}{m_W^2} - 2\right)\,p_\chi\,\cos 2\teff\,\,
\cos\theta\,.
\eeq{theta_dist}
The presence of a term linear in $\cos\theta$ indicates parity violation, and 
measuring the size of this term provides information about the 
effective mixing angle $\teff$.   

An analogous calculation for the leptonically decaying top results in a
matrix element squared proportional to
\beq
 \left | {\cal M} \right |^2 \propto E_\chi + \sin2\teff m_\chi+p_\chi\,\cos 2\teff\,\,
\cos\hat{\theta}_l\,.
\eeq{Mlep}
Note that in both cases the magnitude of the neutralino momentum in the top rest frame is fixed because the neutralino is the final state of an effective two body decay.

\end{appendix}

\end{document}